\documentclass[
 prxquantum,twocolumn,
 amsmath,amssymb,
 aps,nofootinbib,
]{revtex4-2}

\usepackage{amsthm}
\usepackage{graphicx}
\usepackage{color}
\usepackage{dcolumn}
\usepackage{bm}
\usepackage{mathtools}
\usepackage{mathrsfs}
\usepackage{braket}
\usepackage{float}
\usepackage[table]{xcolor}
\usepackage{array}
\usepackage{colortbl}
\usepackage{tabularx}
\usepackage{booktabs}
\usepackage{multirow}
\usepackage{caption}
\usepackage{subcaption}
\usepackage[utf8]{inputenc}

\newcommand{\cA}{\ensuremath{\mathcal{A}}}

\newcommand{\cF}{\ensuremath{\mathcal{F}}}

\newcommand{\cC}{\ensuremath{\mathcal{C}}}
\newcommand{\cN}{\ensuremath{\mathcal{N}}}

\newcommand{\cU}{\ensuremath{\mathcal{U}}}

\newcommand{\cL}{\ensuremath{\mathcal{L}}}

\newcommand{\cM}{\ensuremath{\mathcal{M}}}

\newcommand{\bx}{\ensuremath{\boldsymbol{x}}}

\newcommand{\bg}{\ensuremath{\boldsymbol{g}}}
\newcommand{\bp}{\ensuremath{\boldsymbol{p}}}

\newcommand{\bh}{\ensuremath{\boldsymbol{h}}}

\newcommand{\bs}{\ensuremath{\boldsymbol{s}}}

\newcommand{\bth}{\ensuremath{\boldsymbol{\theta}}}

\newcommand{\bW}{\ensuremath{\boldsymbol{W}}}

\newcommand{\ba}{\ensuremath{\boldsymbol{a}}}
\newcommand{\bb}{\ensuremath{\boldsymbol{b}}}

\newcommand{\bbR}{\ensuremath{\mathbb{R}}}

\newcommand{\kpsi}{\ensuremath{\ket{\psi}}}
\newcommand{\kphi}{\ensuremath{\ket{\phi}}}

\DeclareMathOperator{\tr}{\textup{Tr}}

\DeclareMathAlphabet{\mymathbb}{U}{BOONDOX-ds}{m}{n}

\usepackage{hyperref}
\hypersetup{
    breaklinks = true,
    colorlinks = true,
    linkcolor = {blue},
    citecolor = {blue},
    urlcolor = {blue},
}

\usepackage[whole]{bxcjkjatype}
\usepackage{ulem}

\begin{document}
\title{Iterative Quantum Feature Maps}

\author{Nasa Matsumoto}
\email{matsumoto.nasa@fujitsu.com}
\affiliation{Quantum Laboratory, Fujitsu Research, Fujitsu Limited, Kawasaki, Kanagawa 211-8588, Japan}

\author{Quoc Hoan Tran}
\email{tran.quochoan@fujitsu.com}
\affiliation{Quantum Laboratory, Fujitsu Research, Fujitsu Limited, Kawasaki, Kanagawa 211-8588, Japan}

\author{Koki Chinzei}
\affiliation{Quantum Laboratory, Fujitsu Research, Fujitsu Limited, Kawasaki, Kanagawa 211-8588, Japan}

\author{Yasuhiro Endo}
\affiliation{Quantum Laboratory, Fujitsu Research, Fujitsu Limited, Kawasaki, Kanagawa 211-8588, Japan}

\author{Hirotaka Oshima}
\affiliation{Quantum Laboratory, Fujitsu Research, Fujitsu Limited, Kawasaki, Kanagawa 211-8588, Japan}

\date{\today}

\begin{abstract}
Quantum machine learning models that leverage quantum circuits as quantum feature maps (QFMs) are recognized for their enhanced expressive power in learning tasks. Such models have demonstrated rigorous end-to-end quantum speedups for specific families of classification problems. However, deploying deep QFMs on real quantum hardware remains challenging due to circuit noise and hardware constraints. Additionally, variational quantum algorithms often suffer from computational bottlenecks, particularly in accurate gradient estimation, which significantly increases quantum resource demands during training. We propose Iterative Quantum Feature Maps (IQFMs), a hybrid quantum-classical framework that constructs a deep architecture by iteratively connecting shallow QFMs with classically computed augmentation weights. 
By incorporating contrastive learning and a layer-wise training mechanism, the IQFMs framework effectively reduces quantum runtime and mitigates noise-induced degradation.
In tasks involving noisy quantum data, numerical experiments show that the IQFMs framework outperforms quantum convolutional neural networks, without requiring the optimization of variational quantum parameters. Even for a typical classical image classification benchmark, a carefully designed IQFMs framework achieves performance comparable to that of classical neural networks.
This framework presents a promising path to address current limitations and harness the full potential of quantum-enhanced machine learning.

\end{abstract}

\pacs{Valid PACS appear here}

\maketitle

\section{Introduction}

Transforming classical data into quantum states by leveraging the exponentially large Hilbert space accessible to quantum computers has gained attention in recent years~\cite{halvlicek:2019:supervised,schuld:2019:feature}. 
This approach draws inspiration from classical machine learning (ML), where input data is mapped into a new feature space to enhance separability.
In quantum machine learning (QML), such transformations are termed quantum feature maps (QFMs).
These maps arise from quantum system dynamics, driven by input data and tunable variational parameters. QFMs highlight the potential of QML, suggesting an exponential speedup due to the classical intractability of simulating certain quantum correlations~\cite{bravyi:2018:qadv}. For example, based on the complexity of the discrete logarithm problem, the first suggested exponential advantage in QML was demonstrated through the calculation of a support vector machine kernel matrix on a fault-tolerant quantum computer~\cite{liu:2020:rigorous}. 
Theoretically, QFMs could enable QML models to serve as universal approximators of continuous functions~\cite{tran:2021:prl:uap}.
It is also possible to design specialized data sets that show the significant differences between quantum and classical models from a learning theory perspective, thereby illustrating the quantum advantage in ML problems~\cite{huang:2021:power}. 
However, finding similar advantages in practical applications, particularly those feasible for near-term quantum computers, remains challenging.

\begin{figure*}
		\includegraphics[width=17cm]{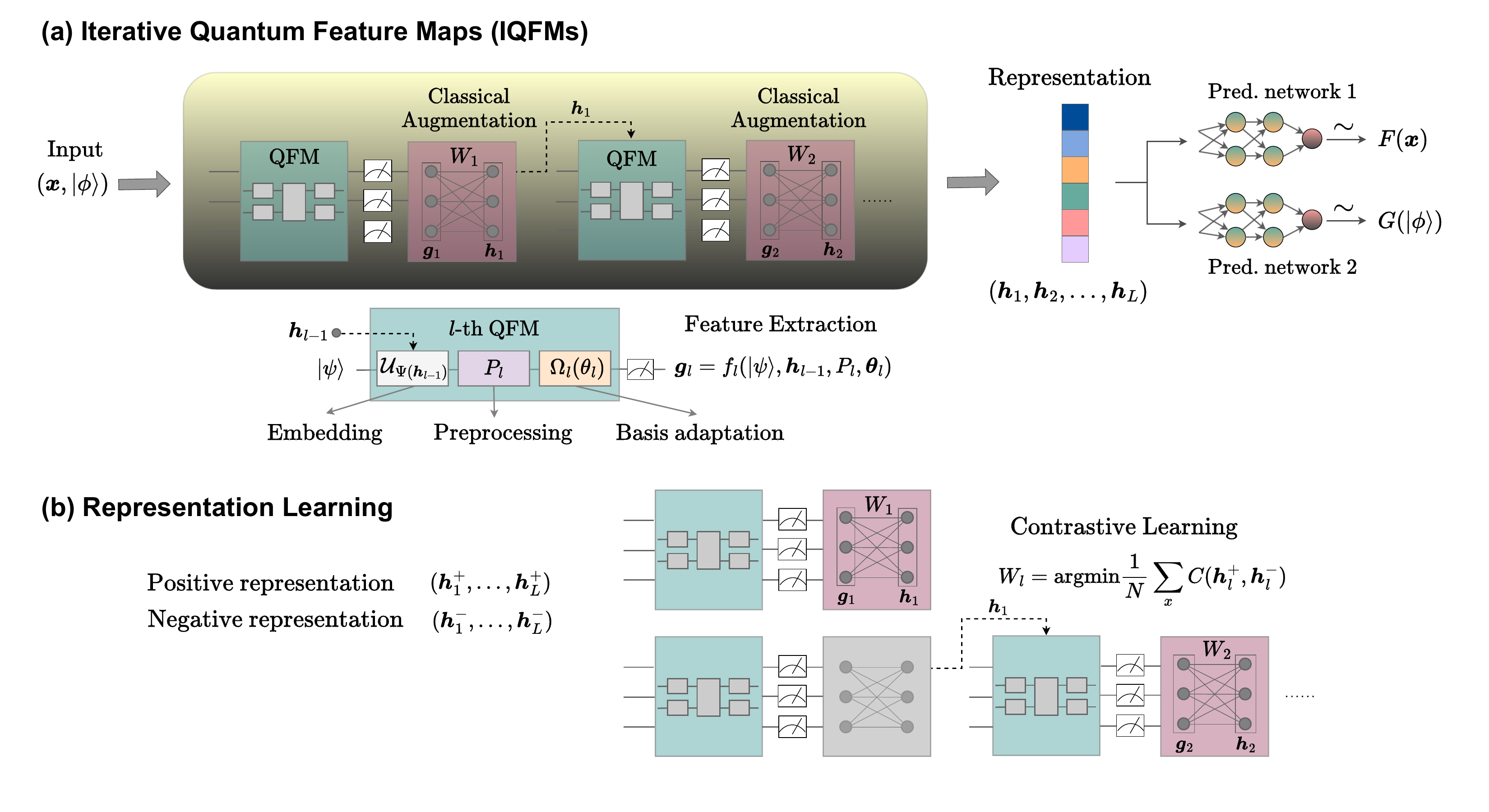}
		\protect\caption{IQFMs and the representation learning in processing classical input $\bx$ and quantum input $\ket{\phi}$. (a) The $l$-th QFM starts with a quantum state $\kpsi$ (re-prepared at each layer) and uses an embedding circuit $\cU_{\Psi(\bh_{l-1})}$ to map classical features $\bh_{l-1}$ into a quantum state, followed by a preprocessing circuit $P_l$, then a parameterized circuit $\Omega_l(\bth_l)$ to adapt the measurement basis. Extracted features $\bg_l$ from the $l$-th QFM are passed to the $(l+1)$-th QFM through classical augmentation with trainable weights. 
        The outputs $\bh_l$ from all augmentation layers are aggregated into a classical feature vector, which is fed into multiple prediction networks to predict properties of $\bx$ and $\ket{\phi}$. 
        At each layer, the quantum circuit is executed on a freshly prepared input state; the post-measurement quantum state is discarded, and only the classical measurement statistics are propagated to the next layer.
  (b) The IQFMs framework employs contrastive learning to train the classical augmentation weights sequentially, leaving quantum circuit parameters fixed.
  Here, positive representations $(\bh_1^{+}\ldots,\bh_L^{+})$ are derived from data similar to the input (e.g., perturbed versions or same-class samples), while negative representations $(\bh_1^{-}\ldots,\bh_L^{-})$ come from dissimilar data (e.g., different-class samples). For each layer $l$, the weight $\bW_l$ is trained to minimize the cost function $\cC(\bh_1^{+},\bh_1^{-})$, which pulls positive representations closer and pushes negative representations away. This process begins with $\bW_1$, fixes it, then trains $\bW_2$, and continues layer by layer until the final layer, then repeating until termination conditions are met.}
		\label{fig:IQFM:overview}
\end{figure*}

The selection of QFMs is often aimed at boosting the quantum model's performance, where typically these maps are neither optimized nor trained. One could use deep quantum circuits to construct a QFM, but this approach does not align well with the capabilities of near-term quantum computers, which are susceptible to noise and prone to errors. A common strategy is to append variational quantum circuits (VQCs) after the feature map circuit, forming the structure of a quantum neural network (QNN). This design aims to create more expressive circuits for quantum algorithms. 
Advancements in quantum embeddings~\cite{lloyd:2020:quantum:embeddings} and neural quantum kernels~\cite{rodgriguez:2025:prr:neural:kernel}, which train problem-inspired kernel functions via QNNs to mitigate exponential concentration and boost generalization, further underscore the value of such hybrid designs in QML.
However, training such circuits~\cite{mitarai:2018:circuit} through variational quantum algorithms (VQAs) often runs into issues like becoming stuck in local minima~\cite{ansachuetz:2022:natcom:VQA,bittel:2021:prl:VQANP} or encountering plateau regions~\cite{clean:2018:natcom:barren,cerezo:2021:natcom:barren,marrero:2021:prxquant:barren}, making it hard to find optimal solutions.

Despite some QNN designs showing resilience to the barren plateaus phenomenon, evidence indicates that the loss functions of these models may be classically simulated if classical data from quantum devices are gathered during an initial phase~\cite{cerezo:2021:natcom:barren,fuster:2024:VQA:dequan,bermejo:2024:qcnn:sim,lerch:2024:quant:sim}. This underscores the importance of developing QML models that are not only highly trainable but also practically useful or capable of achieving quantum advantages beyond classical simulation capabilities. 
Moreover, combined with the lack of efficient gradient estimation algorithms~\cite{Abbas2023-hy, Chinzei2025-pa}, it is argued that VQAs may struggle to surpass classical approaches in terms of time scaling to demonstrate quantum advantages with the current QML models~\cite{liu:2023:vqa:time}. 
This analysis points towards a more natural form of hybrid quantum-classical algorithms that can integrate the strengths of both quantum and classical ML models.

In this paper, we introduce Iterative Quantum Feature Maps (IQFMs), a hybrid quantum-classical framework designed to minimize the quantum resources needed for learning compared to a deep QFM. The IQFMs framework features a deep architecture that iteratively connects shallow quantum circuits as QFMs. This design enhances expressiveness by linking each QFM to the next via measurement outputs, which are then processed through classical augmentation [Fig.~\ref{fig:IQFM:overview}(a)]. QFM outputs are measured across multiple bases, ensuring stability in deeper layers through normalized classical augmentation. The outputs from all augmentation layers are aggregated into a classical feature vector, which is suitable for multiple tasks such as classification or regression using other classical neural networks (NNs) for prediction. Unlike conventional methods that train the parameters of quantum circuits, the IQFMs framework optimizes only the weights of the classical augmentation that links the QFMs to produce effective feature vectors.
By employing contrastive learning~\cite{hadsell:2006:dimensionality,gutmann:2010:contrastive,van:2018:representation,he:2020:momentum,chen:2020:simple} and a layer-by-layer training approach to adjust these classical weights, the IQFMs framework reduces quantum computational runtime and mitigates the impact of noise [Fig.~\ref{fig:IQFM:overview}(b)]. The framework is versatile, supporting both quantum and classical data classification tasks. 
Numerical experiments in our study demonstrate that the IQFMs framework is effective for both quantum data and classical data. In classifying quantum phases of matter using noisy quantum data, the IQFMs framework outperforms the quantum convolutional neural network (QCNN) model ~\cite{cong2019quantum, Chinzei2024-nm, Chinzei2024-tl} without requiring the optimization of variational quantum parameters.
In classifying image data, the IQFMs framework achieves accuracy comparable to that of similar classical architectures.

IQFMs can be understood in the context of the fundamental trade-off between classical simulability, trainability, and quantum advantage. Highly expressive quantum circuits that may resist classical simulation are often difficult to train due to barren plateaus, while shallow or structured circuits that remain trainable tend to be classically simulable. Rather than attempting to overcome this tension by relying on a single deep quantum circuit, IQFMs adopt a different strategy: they iteratively combine shallow quantum feature maps with classical augmentation. This design seeks to improve practical trainability by avoiding variational optimization of quantum parameters, while retaining the potential to leverage measurement-induced features that could become classically challenging in more complex regimes.

\section{Methods}

\subsection{Iterative Quantum Feature Maps (IQFMs)}

The IQFMs framework employs a deep structure that enhances the connectivity of QFMs by applying classical augmentation to the measurement outcomes before moving to the subsequent QFM.
The $l$-th QFM block starts from the input quantum state $\kpsi$, applies a feature map circuit $\cU_{\Psi(\bh_{l-1})}$ to embed the classical data $\bh_{l-1}$ into the quantum state $\ket{\Psi(\bh_{l-1})}$, then uses a fixed preprocessing circuit $P_l$ to entangle and mix the data, and applies measurement basis adaptation circuit $\Omega_l(\bth_l)$ parameterized by $\bth_l$ to rotate the state into an optimal measurement basis.
Here, the subscript $\Psi$ in $\cU_{\Psi}$ labels the chosen feature-map (embedding).
This process yields $\ket{\Psi(\bh_{l-1},P_l,\bth_l)}=\Omega_l(\bth_l) P_l \cU_{\Psi(\bh_{l-1})} \kpsi$~[Fig.~\ref{fig:IQFM:overview}(a)]. 
The sequential repetition of the combination of $\cU_{\Psi(\bh_{l-1})}$, $P_l$, and $\Omega_l(\bth_l)$, with varying parameters $\bth_l$, is known as the data re-uploading framework~\cite{perez:2020:reupload}.
The feature extraction via measurements yields a feature vector $\bg_l=f_l(\kpsi,\bh_{l-1}, P_l,\bth_l)$ $\in \mathbb{R}^{d_g}$, where $d_g$ is the number of extracted features. In general, $f_l(\kpsi,\bh_{l-1}, P_l, \bth_l)$ can be constructed from measurement records. Here, we restrict its implementation to the expectation values of measurement operators, i.e., $[f_l(\kpsi,\bh_{l-1}, P_l, \bth_l)]_j=\braket{\Psi(\bh_{l-1},P_l,\bth_l)| O_j | \Psi(\bh_{l-1},P_l,\bth_l)}$, where $O_j$ is the observable associated with the $j$-th feature.

The IQFMs framework with $L$ layers is constructed as follows:
\begin{enumerate}
    \item At the layer $l=1$, the QFM is initialized with a quantum data $\kpsi$ and a classical data $\bh_0$.
    \item For each layer $l=1,\ldots,L$, the following operations are performed:
         \begin{align}
         \bg_l &= f_{l}(\kpsi, \bh_{l-1}, P_l, \bth_{l}) \in \bbR^{d_g}\label{eqn:fet:def1}, \\
         \bh_l &= \cA_l(\bW_l, \hat{\bg}_l) \in \bbR^{d_h} \label{eqn:fet:def2}.
         \end{align}
\end{enumerate}
Here, Eqs.~\eqref{eqn:fet:def1} and \eqref{eqn:fet:def2} represent the quantum feature extraction via the QFM of the $l$-th layer and the classical augmentation $\cA_l$ with parameters $\bW_l$ (such as a classical neural network), respectively.
In the infinite-shot limit, $\bg_l$ is given by exact expectation values. Under finite measurement statistics, the circuit output is the finite-shot estimator $\hat{\bg}_l$, which is used as the input to the classical augmentation.
In this framework, the post-measurement quantum state is discarded after each layer, and only the classical measurement statistics are propagated to the subsequent layer. Each QFM layer operates on a freshly prepared quantum state $\kpsi$, into which the measurement statistics are encoded via a feature map.
For ease of analysis, we consider the dimensions of $\bg_l$ and $\bh_l$ as $d_g$ and $d_h$, respectively, to be constant across layers.
At each layer $l$, there are two sets of parameters involved: $\bth_l$ for the quantum circuit and $\bW_l$ for the classical augmentation, corresponding to the quantum feature extraction and classical processing steps, respectively. 
This work assumes that $\bW_l$ is trainable while $\bth_l$ is fixed to random values.

The above structure enables IQFMs to process both classical and quantum data, enhancing its versatility and near-term applicability on quantum devices.
In an ML task involving both classical input $\bx$ and quantum input $\kphi$, our QML model needs to approximate a true hybrid function $\cF(\bx, \kphi)$.
This encompasses scenarios with solely classical input or solely quantum input.
For instance, when the IQFMs framework processes only classical data $\bx$, we set $\bh_0 = \bx$ and fix the quantum state $\kpsi$ (e.g., $\ket{0}^n$) to approximate a target function $F(\bx)$. Conversely, when the IQFMs framework processes only the quantum state $\kphi$, we set $\kpsi = \kphi$ and fix the classical data $\bh_0\in \mathbb{R}^{d_h}$ to approximate a target function $G(\kphi)$ (Fig.~\ref{fig:IQFM:overview}).

When inputting a quantum state into IQFMs, a preprocessing circuit $P_l$ plays a crucial role. Feeding $\kpsi$ into the $l$-th QFM without preprocessing may lead to loss of specific features of the quantum state information. The preprocessing circuit enhances the distinguishability of these relevant features before measurement.
Furthermore, the re-input structure can be essential for IQFMs, especially in deep architectures. If the input quantum data $\kphi$ is used only at the first layer, information transmission may become insufficient as the depth increases. The re-input structure addresses this issue by supplying the same input state $\kpsi=\kphi$ to every layer, promoting learning stability in deeper layers.

Remarkably, IQFMs can be defined on a modular architecture, in which the classical data $\bh$ is partitioned into multiple parts and processed in parallel by multiple QFMs within a single module. The resulting feature vectors for each module are then combined, augmented, and forwarded to the QFMs in the next layer.
This modular architecture enables more flexible and expressive models that are capable of capturing complex patterns and relationships in data. The processing pipeline falls under the general definition in Eqs.~\eqref{eqn:fet:def1} and \eqref{eqn:fet:def2}, assuming the embedding, preprocessing, and measurement basis adaptation circuits are disentangled into multiple subcircuits.
Due to this modular design, the entire quantum circuit does not need to be executed on a quantum device at once. Instead, only the individual subcircuits need to be implemented, making it feasible to address large-scale learning tasks even on near-term quantum devices with limited qubit resources.

\subsection{Quantum Feature Extraction}

Quantum feature extraction in the QFM involves performing quantum measurements in multiple bases to extract richer features. We consider a scenario where an experimenter has access to multiple copies of the $n_q$-qubit preprocessed state $\ket{\pi}=P \cU_{\Psi(\bh)} \ket{\psi}$ of the QFM. The experimenter can perform a limited set of quantum measurements across multiple bases, denoted by a collection of Positive Operator-Valued Measurements (POVMs), $\{\cM_k\}$, where each $\cM_k=\{M_{k,j}\}$ corresponds to a specific measurement basis labeled by $k$. 
Here, a POVM is a mathematical tool used in quantum mechanics to describe the possible outcomes of a measurement. 
The operators $M_{k,j}$ act on the system's Hilbert space and satisfy the normalization condition $\sum_j M_{k,j} = I$ for each $k$, ensuring that the probabilities of all possible outcomes sum to 1.
For a given POVM $\cM_k$, the experimenter measures a subset of the copies of $\ket{\pi}$ in the corresponding basis. 
The probability of obtaining outcome $j$ in basis $k$ is given by the Born rule: $p_k(j)=\tr(\ket{\pi}\bra{\pi}M_{k,j})$.
Given an observable $O^{(k)}$ associated with the measurement outcomes in the basis $k$, the expectation value is computed as:
\begin{align}
    \braket{O^{(k)}} = \sum_j o_{k,j} p_k(j),
\end{align}
where $o_{k,j}$ is the numerical value of the observable $O^{(k)}$ corresponding to outcome $j$ in basis $k$.
The experimenter repeats this measurement process over many copies of $\ket{\pi}$ to estimate the probabilities $p_k(j)$ and compute $\braket{O^{(k)}}$.

In practice, the expectation value $\braket{O^{(k)}}$ cannot be accessed directly and is estimated from a finite number of measurement shots. Let $N_k$ denote the number of shots performed for the POVM $M_k$, and let $n_{k,j}$ be the number of occurrences of outcome $j$, satisfying $\sum_j n_{k,j}=N_k$. The outcome probabilities are estimated as $\hat{p}_k(j)=n_{k,j}/N_k$, which yields the finite-shot estimator $\widehat{\braket{O^{(k)}}}=\sum_j o_{k,j} \hat{p}_k(j)$.

Given a set of observables $\{ O_1^{(k)}, \ldots, O_F^{(k)}\}$ associated with the measurement bases, the experimenter collects statistics from each basis $k$, such as the expectation values $g^{(k)}_{i}=\braket{O_i^{(k)}}$, over the index $i\in\{1,\ldots,F\}$.
These values form a feature vector $\bg^{(k)} = (g_1^{(k)}, \ldots, g_F^{(k)} ) \in \bbR^{F}$. 
Employing multiple measurement bases can distinguish quantum and classical generative models~\cite{gao:2022:prx:corr,rudolph:2022:prx:gen} and may enhance QFM's classification performance. This approach also prevents classical simulation of Pauli-Z words expectation values for specific circuit classes~\cite{nest:2011:expectation}.

\begin{figure}
		\includegraphics[width=8cm]{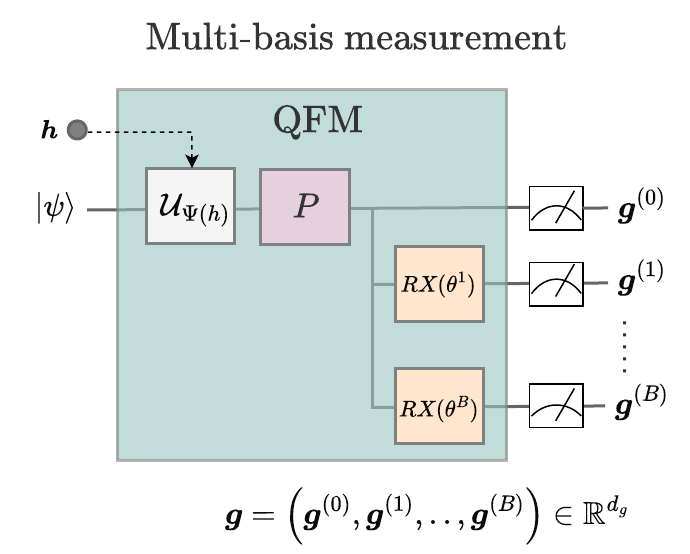}
		\protect\caption{The quantum feature extraction involves multi-basis measurement. The initial feature vector $\bg^{(0)}$ is derived from all qubits measurement in the Z-basis. To expand the feature set, measurements are performed on other $B$ distinct bases to acquire $\bg^{(1)}, \ldots, \bg^{(B)}$. Here, $\bg^{(i)}$ is obtained by applying a random single-qubit rotation $RX(\theta^i)$ to all qubits. 
        The total extracted feature vector $\bg$ is a concatenation of $\bg^{(0)}, \ldots, \bg^{(B)}$.
  \label{fig:feature}}
\end{figure}

We define basis $k=0$ as the computational basis, where each $O_i^{(0)} = Z^{i_1}\otimes \ldots \otimes Z^{i_{n_q}}$ ($\{i_1,\ldots,i_{n_q}\} \in \{0,1\}^{n_q}$) is a Pauli-Z word chosen from $2^{n_q}$ possible Pauli-Z words.
Here, the measurement is performed on $\ket{\pi}$ without the basis adaptation circuit $\Omega(\bth)$.
To enrich the feature set, we perform measurements in additional bases, each indexed by $k$, yielding a feature vector $\bg^{(k)}$. This is accomplished by applying a well-selected  $\Omega(\bth^k)$ to $\ket{\pi}$ before measuring it in a computational basis, corresponding to $O_i^{(k)}=\Omega^\dag(\bth^k) O_i^{(0)} \Omega(\bth^k)$. The circuit $\Omega(\bth^k)$ can be a random Clifford or a random local circuit, such as layers of random single-qubit rotations and two-qubit gates on neighboring qubits. 
To prioritize practicality, we implement $\Omega(\bth^k)$ as a product of single-qubit rotations, $\Omega(\bth^k)=\prod_{j=1}^{n_q}RX(\theta^k_j)$, where $\theta^k_j$ is a randomly fixed angle. The feature for each QFM is a concatenation of $\bg^{(0)}$ with $\bg^{(1)}, \ldots, \bg^{(B)}$, derived from other $B$ distinct bases (Fig.~\ref{fig:feature}): $\bg=(\bg^{(0)},\bg^{(1)},\cdots,\bg^{(B)}) \in \mathbb{R}^{d_g}$, where $d_g=(B+1)F$.

In IQFMs, the final representation vector is formed by augmenting $\bg_l$ as $\bh_l=\cA(\bW_l,\bg_l) \in \bbR^{d_h}$ with the classical weights $\bW_l$ and concatenating them from all layers into $\bh=(\bh_1,\cdots,\bh_L) \in \bbR^{d_hL}$.
This representation vector $\bh$ can be utilized for downstream tasks such as classification and regression. 
For example, a trainable classical neural network $\cN$ can process $\bh$ to approximate the target functions, e.g., $\cN(\bh)\sim F(\bx)$ or $\cN(\bh)\sim G(\ket{\phi})$ [Fig.~\ref{fig:IQFM:overview}(a)].
Numerical experiments in our study reveal that, in most cases, only training $\cN$ without learning $\bW_l$ cannot reach sufficient accuracy in classification tasks. This demonstrates that further representation learning is required to enhance the separability of $\bh$ across distinct input labels.

While multi-basis measurements enrich the extracted feature space, the design of the underlying circuits must also account for sampling-related scalability issues. In particular, when the fixed circuits produce highly entangled states, expectation values of observables can exhibit exponential concentration around input-independent values, which can render the extracted features effectively input-independent unless an exponentially large number of measurement shots is used. Such limitations have been discussed in the context of QELM and quantum reservoir computing~\cite{xiong:2025:fundamental,sannia:2025:concentration}. In IQFMs, we mitigate this risk in practice by using shallow circuits, local measurements, and a modular architecture with small quantum subsystems.

\subsection{Representation Learning}

Quantum circuit learning~\cite{mitarai:2018:circuit} focuses on optimizing the parameters of quantum circuits, whereas IQFMs training targets the parameters of classical augmentation $\bW_l$ without adjusting the quantum circuit parameters.
However, training these classical parameters using a conventional approach, such as defining a loss function based on the final output of IQFMs and the target of ML tasks and then updating the parameters via its gradient, still necessitates computing the gradient of the QFM output. This is a computationally expensive process that we aim to avoid.
To estimate the quantum resources required for this training, we fix the size of the training set, the number of measurement shots per feature extraction, and the per‐layer circuit runtime. Let $T$ denote the number of training epochs and consider the loss function $\cL = \cL(\bh_1, \ldots, \bh_L)$,
where each augmented feature $\bh_l \in \bbR^{d_h}$ is produced in the $l$-th layer. The gradient with respect to $\bW_l$ can be written as
\begin{align}
\frac{\partial \cL}{\partial \bW_l}
&= \sum_{k=l}^L
\frac{\partial \cL}{\partial \bh_k}
\,\frac{\partial \bh_k}{\partial \bW_l} \nonumber\\
&= \sum_{k=l}^L
\frac{\partial \cL}{\partial \bh_k}
\Biggl(\prod_{m=l+1}^k \frac{\partial \bh_m}{\partial \bh_{m-1}}\Biggr)
\frac{\partial \bh_l}{\partial \bW_l} \nonumber\\
&= \sum_{k=l}^L
\frac{\partial \cL}{\partial \bh_k}
\Biggl(\prod_{m=l+1}^k \frac{\partial \bh_m}{\partial \bg_m}
\;\frac{\partial \bg_m}{\partial \bh_{m-1}}\Biggr)
\frac{\partial \bh_l}{\partial \bW_l}\,,
\end{align}
where $\bg_m \in \bbR^{d_g}$ is the quantum‐extracted feature vector at layer $m$. The terms $\partial \cL/\partial \bh_k$, $\partial \bh_m/\partial \bg_m$, and $\partial \bh_l/\partial \bW_l$ can be computed classically, so the only quantum‐costly Jacobian is $\partial \bg_m/\partial \bh_{m-1}$. Using the parameter-shift rule~\cite{mitarai:2018:circuit,schuld:2019:shift}, evaluating this $d_g\times d_h$ Jacobian requires $O(d_h\,d_g)$ circuit executions per layer. Summing over $L$ layers and $T$ training epochs yields a total quantum‐resource scaling of
$O\bigl(T\,L\,d_h\,d_g\bigr)\,.$

Alternatively, IQFMs can be employed without training any parameters, transforming it into a quantum extreme learning machine (QELM)~\cite{huang:2006:elm,mujal:2021:opportunities}, although this may introduce some drawbacks.
In the QELM, internal parameters are typically randomized, and the extracted features are fed into a trainable readout layer for classification or regression tasks. 
The rationale for using QELM lies in leveraging quantum dynamics as a quantum reservoir~\cite{fujii:2017:qrc,nakajima:2019:qrc,tran:2021:temporal,kubota:2023:temporal} to generate a high-dimensional feature space suitable for ML tasks, bypassing the challenges of training internal dynamics. The classical augmentation can be viewed as a higher-order connection~\cite{tran:2020:higherorder} that enhances the computational power of quantum reservoirs. 
As internal parameters require no training, quantum resources are utilized solely for feature extraction, resulting in a computational cost of $O(Ld_g)$.
However, this approach is sensitive to the dynamical regime, lacks robustness against noise, and may fail to ensure input separability, as evidenced by our numerical experiments.

To address the challenge of using IQFMs as QELM, we propose a representation learning algorithm for IQFMs to obtain extracted features with high input separability.
In the applications, we perform two steps: Step 1 involves representation learning up to $T$ epochs to refine the extracted features, while Step 2 trains the readout layer for classification or regression tasks using these features.

The representation learning leverages contrastive learning and a layer-by-layer training strategy to optimize its classical augmentation components [Fig.~\ref{fig:IQFM:overview}(b)]. 
The key idea of IQFMs lies in training the classical augmentation components during representation learning without computing the gradient of the QFM output. 
This scheme does not remove the optimization of internal parameters as QELM but eliminates the dependency of quantum resources on the number of parameters during optimization, reducing the quantum computational runtime to  \(O(TLd_g) \).

Contrastive learning~\cite{hadsell:2006:dimensionality,gutmann:2010:contrastive,van:2018:representation,he:2020:momentum,chen:2020:simple}, a well-established ML technique, constructs an embedding space where similar sample pairs are positioned closely together, while dissimilar pairs are separated by greater distances. This method enhances the ability to discern meaningful patterns in the data by emphasizing relative relationships between samples rather than absolute values. In the context of IQFMs, contrastive learning strengthens robustness to noise, which is crucial in quantum systems, by focusing on essential similarities and differences rather than noise-induced fluctuations. For example, contrastive learning can mitigate the variability due to quantum measurements or excessive unitary rotations perturbed by hardware imperfections. It can prioritize stable, intrinsic features over transient errors, thereby stabilizing feature extraction across noisy quantum circuits.
The details of contrastive learning are provided in Appendix C.

The layer-by-layer training approach complements the contrastive learning by sequentially optimizing the classical augmentation weights for each QFM layer.
Rather than jointly training all parameters simultaneously using much more quantum resources for optimization, quantum resources are required only for data collection in our method. This strategy reduces computational complexity and avoids the pitfalls of gradient-based optimization in deep quantum circuits, such as barren plateaus~\cite{clean:2018:natcom:barren,cerezo:2021:natcom:barren,marrero:2021:prxquant:barren}, which are prevalent in VQAs. 

We employ a supervised contrastive‐learning protocol at each IQFMs layer $l$ to leverage label information during training. Given a training example $\bx$ (or $\ket{\phi}$) with true label $y$, we first compute its layer‑$l$ augmented feature vector $\bp_l = \bh_l$,
which serves as the anchor. We then sample a positive sample $\bx^+$ (or $\ket{\phi^+}$) sharing the same label $y$, and a negative sample $\bx^-$ (or $\ket{\phi^-}$) with a different label. Passing $\bx^+$ and $\bx^-$ through the first‑layer QFM yields features $\bg_1^+$ and $\bg_1^-$, which are then classically augmented by $\bW_1$ to produce
\begin{align}
\bh_1^+ = \cA(\bW_1, \bg_1^+),
\qquad
\bh_1^- = \cA(\bW_1, \bg_1^-).    
\end{align}
We optimize $\bW_1$ by minimizing a contrastive loss
$\cC\bigl(\bh_1^+, \bh_1^-\bigr),$
thereby encouraging anchor–positive pairs to be closer than anchor–negative pairs in feature space. After convergence at layer 1, we fix $\bW_1$ and repeat the same procedure at layer 2 (training $\bW_2$ with new anchor, positive, and negative samples), and so on through layer $L$. This layer‑wise training continues until a predefined criterion is met, enabling each $\bW_l$ to learn representations that maximize inter‑class separability.

We utilize noise contrastive estimation~\cite{gutmann:2010:contrastive} to address the minimization problem:
\begin{align}\label{eqn:loss:func}
    {\bW_l} = \textup{argmin}\dfrac{1}{N}\sum_{\bx}\cC\left(\bh_l^{+}, \bh_l^{-}\right),
\end{align}
where the cost function $\cC\bigl(\bh_l^{+},\bh_l^{-}\bigr)$ is defined as:
\begin{align}\label{eqn:contrast:loss}
\cC\bigl(\bh_l^{+},\bh_l^{-}\bigr)
&=
\log\!\biggl[1 + \exp\!\Bigl(\frac{\text{cs}(\bh_l^{-}, \bp_l) - \text{cs}(\bh_l^{+}, \bp_l)}{\tau}\Bigr)\biggr].
\end{align}
Here, $\tau$ is the scale parameter, and $\text{cs}$ denotes the cosine similarity function,
$\text{cs}(\ba, \bb)
= \frac{\ba \cdot \bb}{\|\ba\|\,\|\bb\|},
$
where $\ba \cdot \bb$ is the dot product and $\|\ba\|$, $\|\bb\|$ are the vector magnitudes. A similar form of Eq.~\eqref{eqn:contrast:loss} was employed in Ref.~\cite{momeni:2023:science} for training deep physical neural networks. 
This cost is optimized using, for example, gradient descent for $\bW_l$ to bring the augmented vectors $\bh_l^{+}$ and $\bp_l$ closer together, while pushing $\bh_l^{-}$ and $\bp_l$ further apart. 
In the testing phase, the optimized IQFMs and classifier are applied sequentially to test data to predict labels.

In the representation learning, two types of training epochs are defined: outer training epochs and inner training epochs. During each outer epoch, the training samples are shuffled, and positive and negative pairs are selected for contrastive learning. Subsequently, inner epochs are performed to optimize the contrastive cost $\cC\bigl(\bh_l^{+},\bh_l^{-}\bigr)$.
In all numerical experiments in our study, the scale parameter $\tau$ is fixed at 8. The number of outer epochs is set to 100 for quantum data classification and 200 for classical data classification, while each outer epoch comprises 40 inner epochs.

\section{Results}

In this section, we numerically illustrate the effectiveness of IQFMs in classification tasks for both quantum data and classical data.

\subsection{Quantum Data Classification}

We perform two types of quantum phase recognition tasks to verify the versatility of IQFMs.
These tasks aim to classify the ground states of Hamiltonians into several quantum phases.
Quantum phase recognition is not only a standard benchmark in QML but also has potential applications in the field of condensed matter physics~\cite{Wu2024-iy}.

\begin{figure}
          \includegraphics[width=8.7cm]{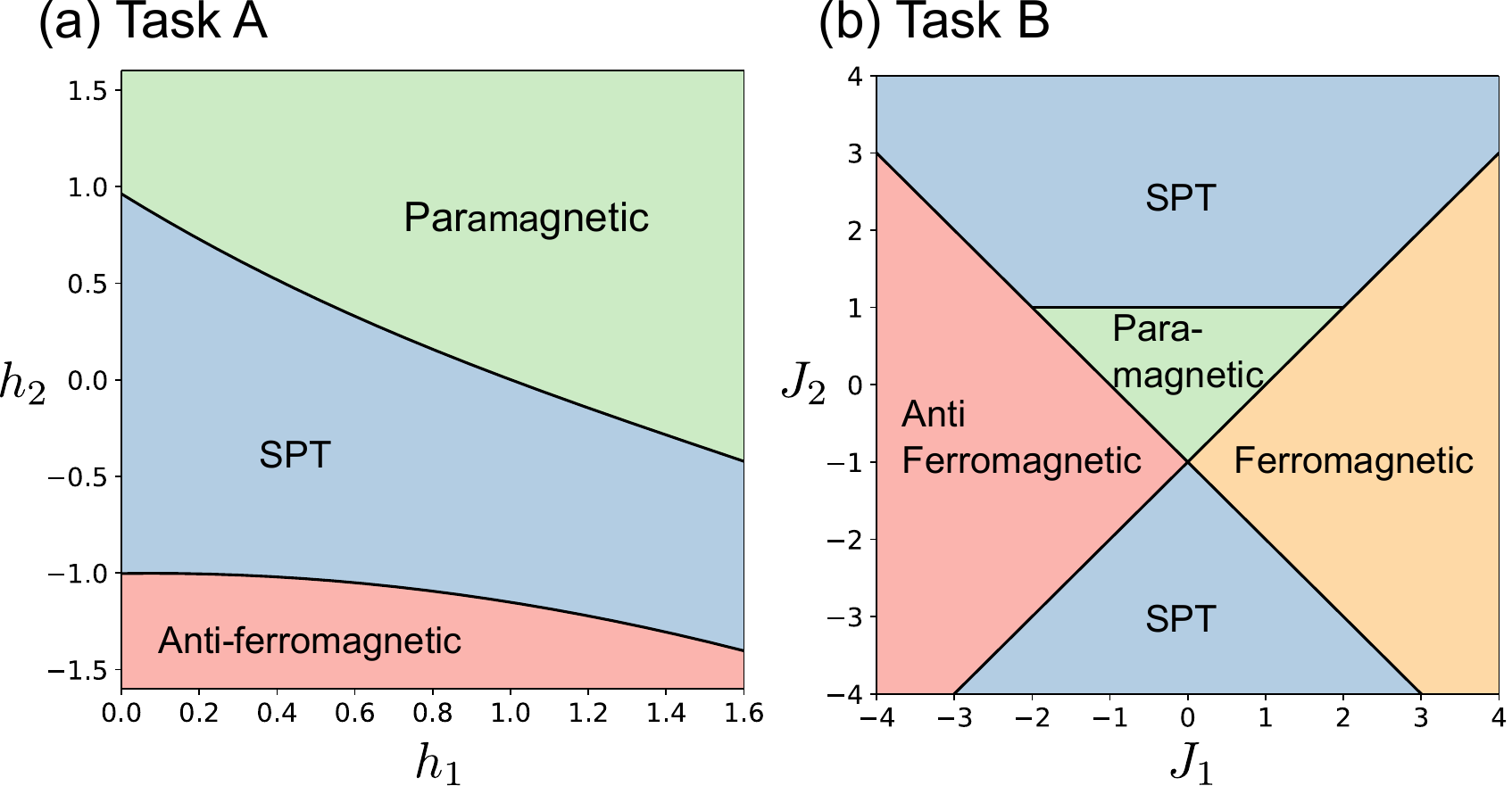}
          \protect\caption{Ground-state phase diagrams of Tasks A and B.\label{fig:quantum:diagram}}
\end{figure}

In the first task (Task A), we consider the following Hamiltonian on the one-dimensional $n$-qubit system with the periodic boundary condition~\cite{cong2019quantum}:
\begin{align}\label{eqn:taska}
H_A = -\sum_{i=1}^{n} Z_i X_{i+1} Z_{i+2} - h_1 \sum_{i=1}^{n} X_i - h_2 \sum_{i=1}^{n} X_i X_{i+1},
\end{align}
where \( h_1 \) and \( h_2 \) are constants, and \( Z_i \) and \( X_i \) are the Pauli matrices at the $i$-th qubit. 
In this Hamiltonian, the competition of interactions gives rise to three distinct quantum phases: the symmetry-protected topological (SPT), paramagnetic, and anti-ferromagnetic phases [Fig.~\ref{fig:quantum:diagram}(a)].
Task A is a binary classification problem distinguishing the SPT phase from the other two phases. 
For training, we use 80 ground state wave functions. 
These are generated by sampling 40 evenly spaced values of $h_1$ in the range [0, 1.6] for each of $h_2 = 0$ and $h_2 = -1.109$. 
For testing, we use 800 test data points obtained by sampling 40 evenly spaced values of $h_1$ in the range [0, 1.6] for each of the following $h_2$ values: $-1.35$, $-1.285$, $-1.225$, $-1.154$, $-1.109$, $-1.079$, $-1.049$, $-1.024$, $-1.0009$, $-1.004$, $-0.3531$, $-0.2479$, $-0.1377$, $-0.02755$, $0.09766$, $0.2229$, $0.3631$, $0.5033$, $0.6636$, $0.8439$.
Note that both the training and test datasets contain data for all three phases.
We use a relatively small training dataset to reflect a data-limited regime often encountered in quantum data settings. This situation arises whether one prepares quantum states on actual hardware or generates them via classical simulation, where preparing and labeling quantum states is typically more restrictive than in classical benchmarks. The test dataset is chosen larger to evaluate generalization across a broad region of the phase diagram.

\begin{figure*}
	\includegraphics[width=17.5cm]{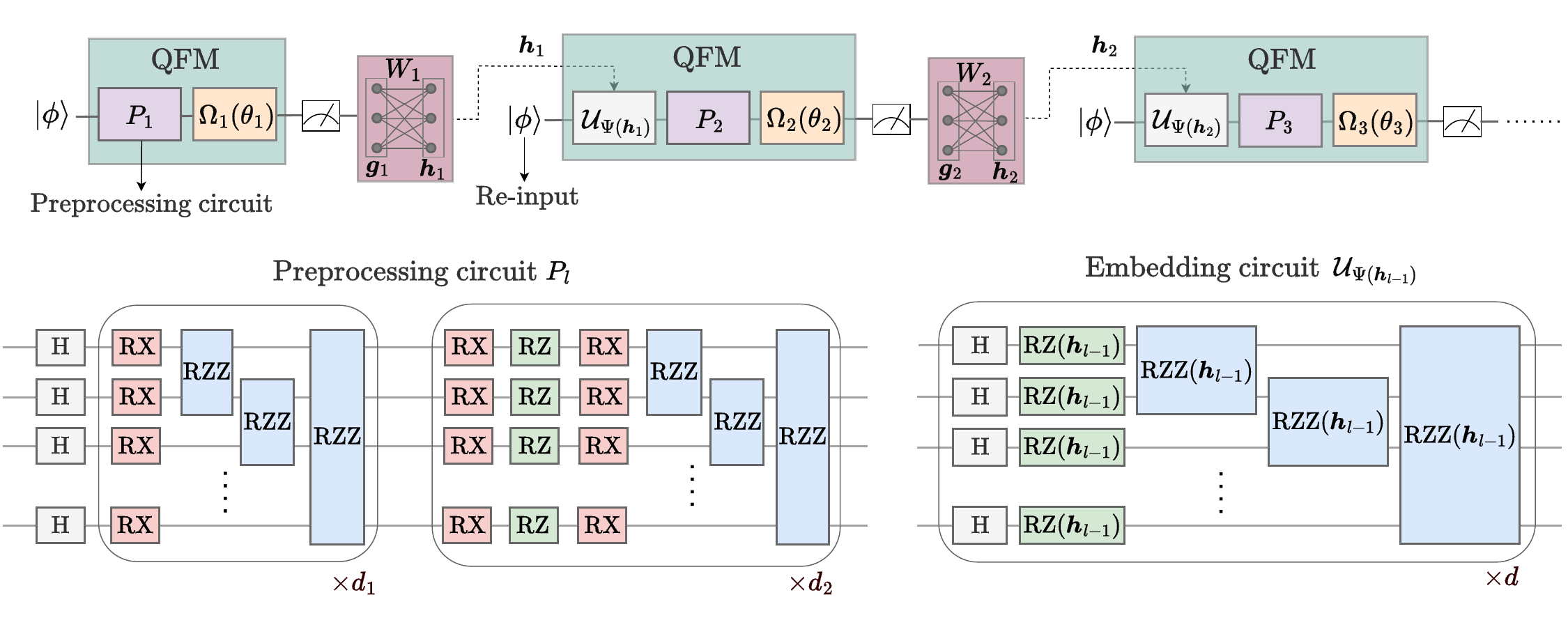}
	\protect\caption{For quantum data classification tasks, the IQFMs framework incorporates a re-input structure: $\kpsi=\kphi$, where $\kphi$ is the input quantum state. Each layer is executed on an independently prepared instance of $\kpsi$. The $l$-th preprocessing circuit $P_l$ starts with Hadamard gates applied to all qubits, followed by several layers of randomly parameterised RX, RZ, and RZZ gates. Starting from the second QFM layer, an embedding circuit $\cU_{\Psi(\bh_{l-1})}$ maps classical data $\bh_{l-1}$ into a quantum state. This circuit comprises $d$ layers, each consisting of Hadamard gates on all qubits, followed by RZ and RZZ gates parameterized by $\bh_{l-1}$. In both the preprocessing and embedding circuits, RX and RZ gates are applied to each qubit individually, and RZZ gates are applied to all adjacent qubit pairs arranged in a circular topology.}
	\label{fig:IQFM:quantum_data}
\end{figure*}

In the second task (Task B), we consider another one-dimensional Hamiltonian with the periodic boundary condition~\cite{PhysRevB.96.165124,PhysRevB.84.165139}:
\begin{align}\label{eqn:taskb}
H_B = \sum_{i=1}^{n} Z_i - J_1 \sum_{i=1}^{n}X_i X_{i+1} - J_2 \sum_{i=1}^{n}X_{i-1} Z_i X_{i+1},
\end{align}
where \( J_1 \) and \( J_2 \) are the coupling strengths. 
The phase diagram of this Hamiltonian includes the SPT, ferromagnetic, anti-ferromagnetic, and paramagnetic phases [Fig.~\ref{fig:quantum:diagram}(b)]. 
Task B is a four-class classification problem distinguishing all these quantum phases.
We sample 50 and 1,000 ground state wave functions for training and test datasets, respectively. For each dataset, the coupling constants $J_1$ and $J_2$ are randomly drawn from the range $[-4, 4]$ and then fixed. We ensure an equal number of data for each class.
These Hamiltonians are chosen to explore different quantum phase transitions and to evaluate the effectiveness of our proposed method in recognizing quantum phases.
As in Task A, the choice of 50 training and 1,000 test datasets reflects a practical trade-off between computational cost and statistical reliability. Class balance is enforced in both datasets to avoid bias across different quantum phases.
We fix $n=8$ and obtain the ground state wave functions by numerically diagonalizing the Hamiltonian matrices defined in Eqs.~\eqref{eqn:taska} and \eqref{eqn:taskb}. 
These ground states are considered input states for IQFMs to predict their corresponding quantum phases. We generated these data based on Ref.~\cite{recio2024learning}.

The structure of IQFMs for this task is described in Fig.~\ref{fig:IQFM:quantum_data}, where the embedding circuit $\cU_{\Psi(\bh_{l-1})}$ and the preprocessing circuit $P_l$ in each QFM layer consist of Hadamard and single- and two-qubit rotation gates.
The embedding circuit is applied only from the second QFM layer onwards.
After applying the embedding and preprocessing circuits, we extract the quantum features $\bg_l$ associated with $n$ observables, $O_i=Z_i$ ($i=1,\cdots,n$), on four different measurement bases.
That is, the dimension of the feature vector obtained in this process is $d_g=4n$.
Then, we classically augment $\bg_\ell$ as $\bh_l=\pi\boldsymbol{\tanh}{(\bW_l\bg_l/2)}\in\mathbb{R}^{d_g}$ and embed it into the next QFM layer, where $\bW_\ell$ is a $d_g\times d_g$ matrix, and $\boldsymbol{\tanh}$ denotes the element-wise hyperbolic tangent function (note that $d_g=d_h$). 
After obtaining $\bh=(\bh_1,\cdots,\bh_L)$ from the representation learning, we train a three-layer feedforward NN to classify them into target phases.
The numerical details of IQFMs are provided in Appendix D.

We compare the performance of IQFMs with the QCNN~\cite{cong2019quantum, Chinzei2024-nm, Chinzei2024-tl}, which is a well-established QML model for these tasks. The QCNN consists of alternating convolutional and pooling layers. The convolutional layers utilize single- and two-qubit rotation gates with variational parameters optimized during training to extract local features. The pooling layers reduce the number of active qubits by half until only one qubit remains. Finally, the expectation value of the Pauli-Z operator is measured to generate output for the classification task.  The details of QCNN are provided in Appendix E. We use the TorchQuantum~\cite{hanruiwang:2022:quantumnas} library in our numerical experiments.

\begin{figure}
		\includegraphics[width=8.7cm]{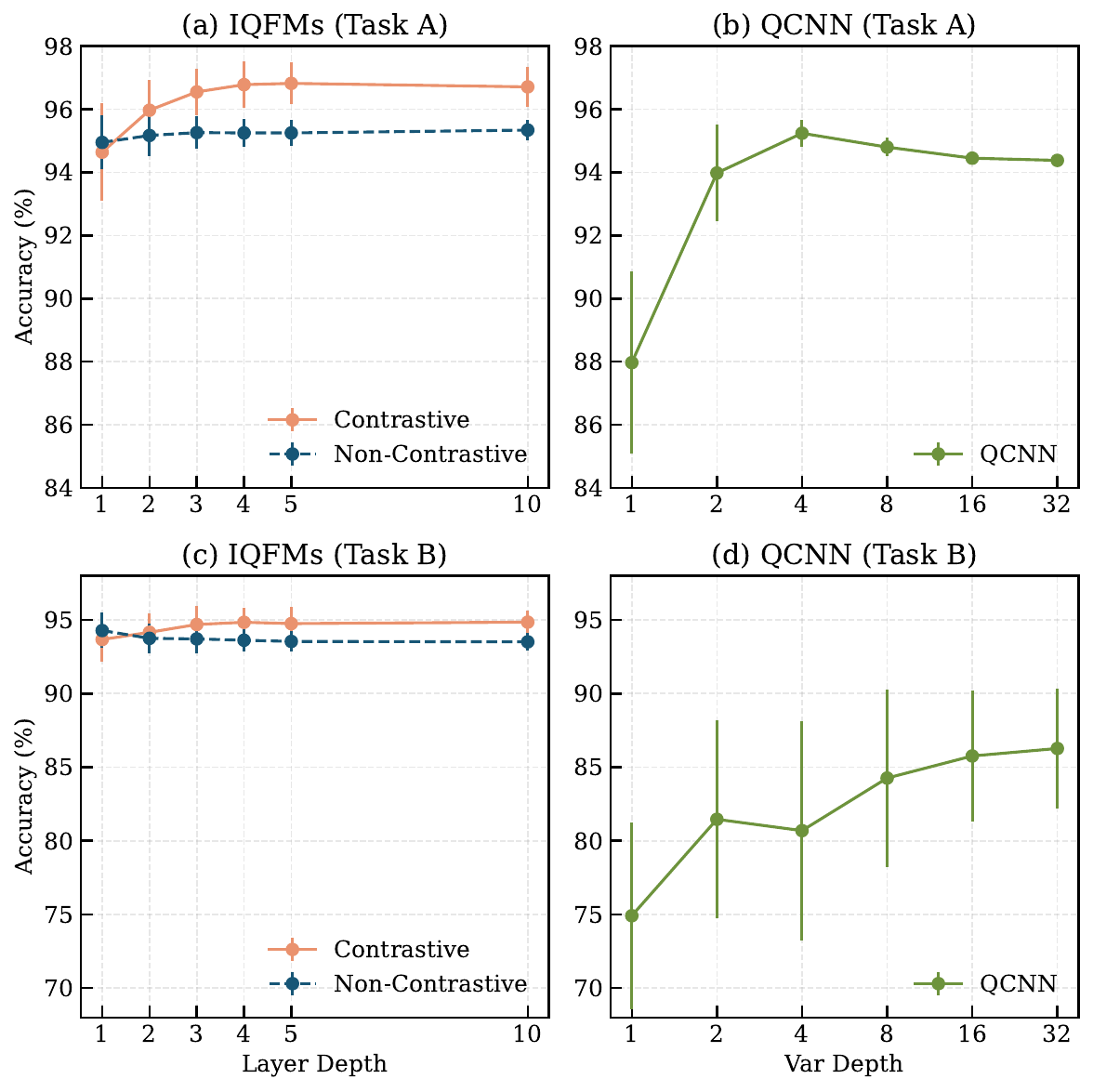}
          \protect\caption{Test accuracy for Tasks A and B using IQFMs and QCNN under different configurations. The solid orange lines, dashed teal-blue lines, and solid green lines represent the average accuracy over 50 trials (with error bars) for IQFMs with contrastive learning, IQFMs with non-contrastive learning, and QCNN, respectively. \label {fig:quantum:taskab}}
\end{figure}

\begin{figure}
		\includegraphics[width=8.7cm]{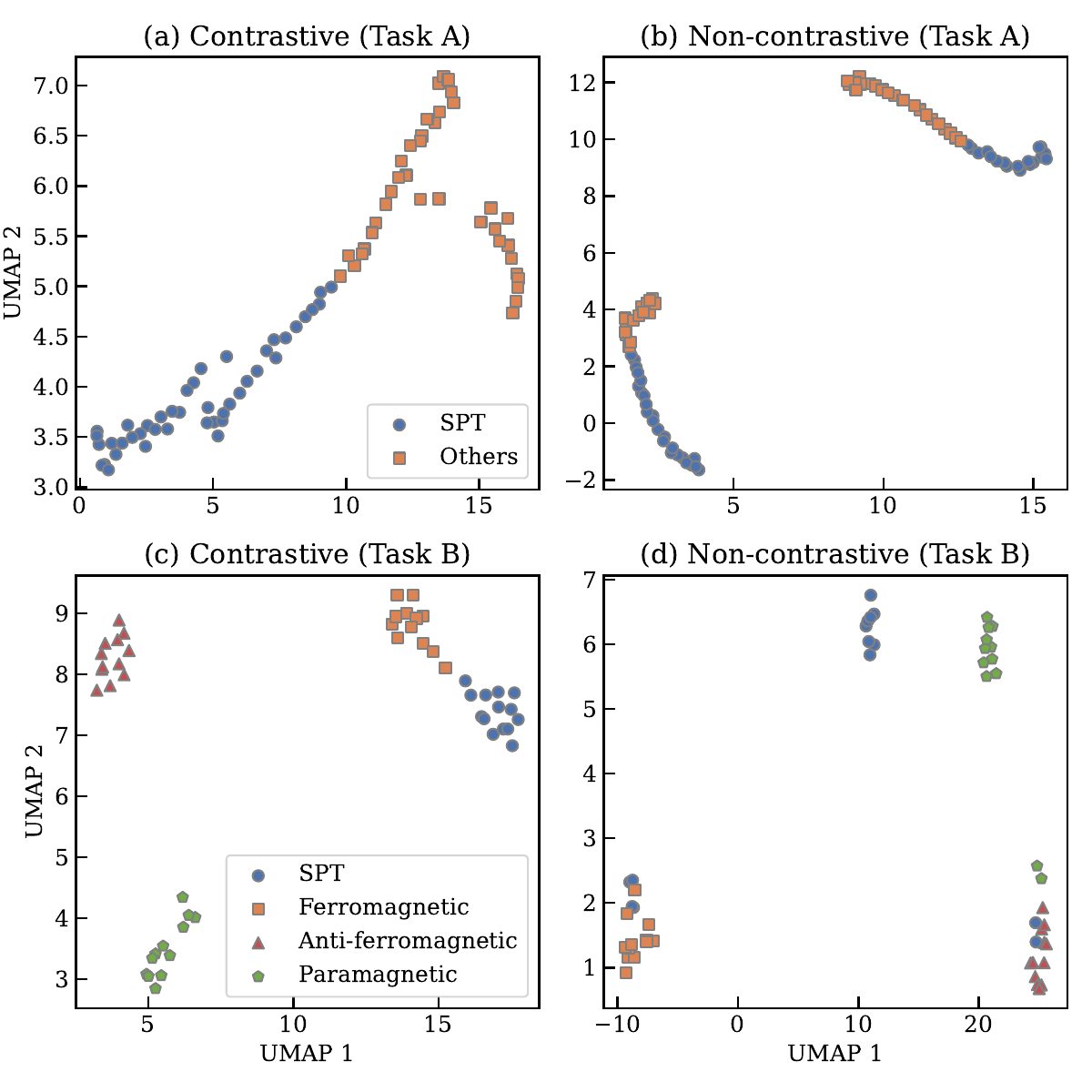}
          \protect\caption{Visualization of the final representation vectors from IQFMs using UMAP~\cite{mcinnes2018umap} for Tasks A and B. Each point represents a test sample, with a marker indicating the corresponding quantum phase.\label{fig:quantum:umap}}
\end{figure}

Figure~\ref{fig:quantum:taskab} presents the test accuracy of IQFMs compared to QCNN, averaged over 50 experimental runs with different random initializations.
Figures~\ref{fig:quantum:taskab} (a) and (c) show the performance of IQFMs across various layer depths $L$, both with (contrastive) and without (non-contrastive) training the classical augmentation weights $\bW_l$.
In non-contrastive case, the weights remain randomly initialized and untrained.
Figures~\ref{fig:quantum:taskab} (b) and (d) show the performance of QCNN for different convolutional layer depths, denoted as var\_depth.
Here, the expectation values of observables are calculated exactly, corresponding to an infinite number of measurement shots.
For both tasks, the IQFMs framework consistently outperforms QCNN in terms of test accuracy, regardless of whether contrastive learning is used, even though the QFM circuits themselves are not optimized. These results suggest that quantum feature extraction via random measurement bases combined with classical post-processing is sufficiently powerful to learn the relevant quantum phases.

Figure~\ref{fig:quantum:taskab} also highlights the benefits of contrastive learning for IQFMs. Training $\bW_l$ with contrastive learning improves the separability of quantum data associated with different quantum phases, leading to higher accuracy across most layer depths $L$.
To further validate the effectiveness of contrastive learning instead of the post processing NN, we visualize the final representation vector $(\bh_1,\cdots,\bh_L) \in \bbR^{d_hL}$ from IQFMs with $L=5$ layers using Uniform Manifold Approximation and Projection (UMAP)  technique~\cite{mcinnes2018umap} for both contrastive and non-contrastive cases, as shown in two-dimensional plots in Fig.~\ref{fig:quantum:umap}. 
The results demonstrate that contrastive learning leads to more clearly separated representations in the representation space, indicating improved discriminability.

\begin{figure}
		\includegraphics[width=8.7cm]{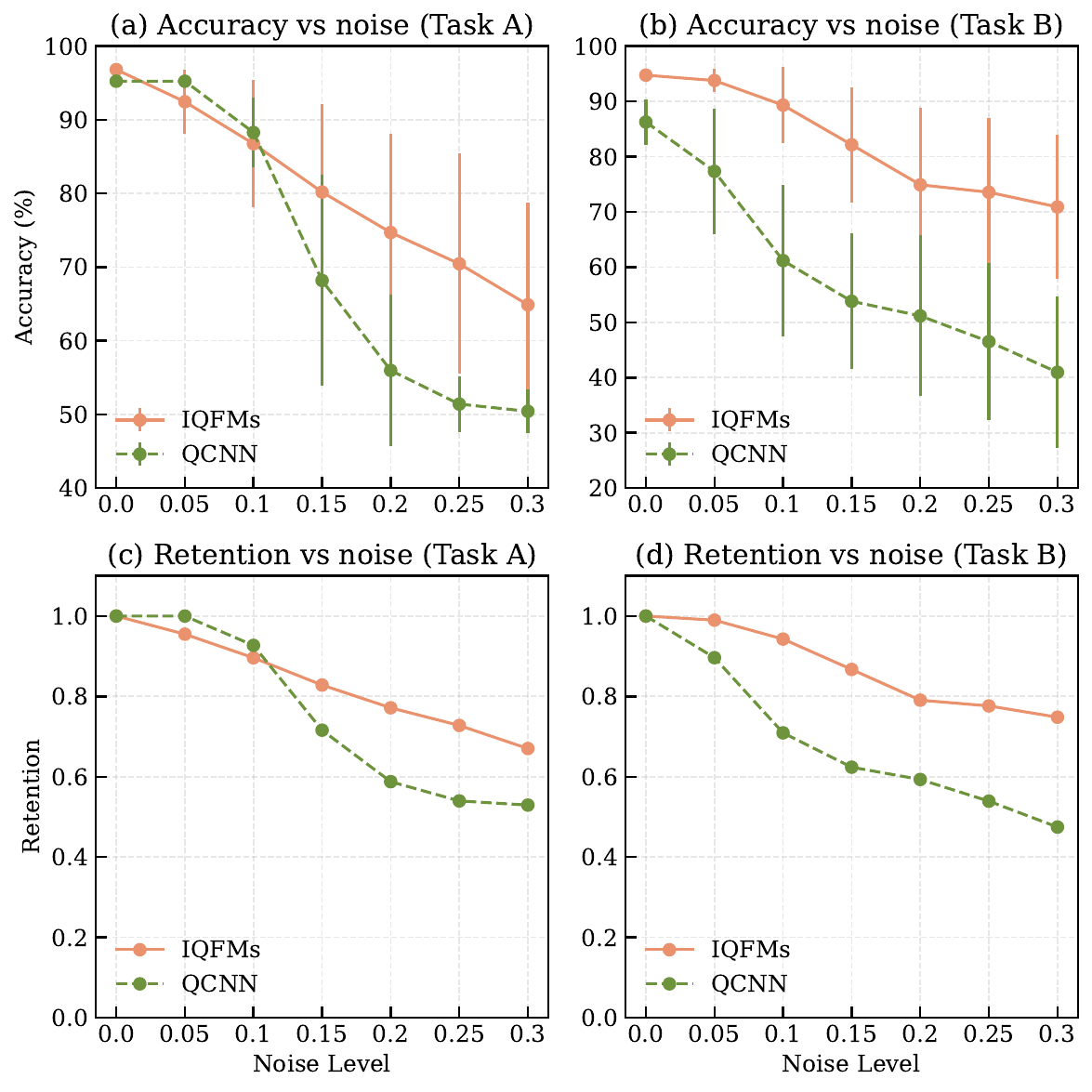}
          \protect\caption{Performance of IQFMs and QCNN across different noise levels applied to the input data for Task A and Task B. (a,b) Test accuracy for Task A and Task B, respectively. The solid orange lines and dashed green lines represent the average accuracy over 50 trials (with error bars) for IQFMs (with contrastive learning) and QCNN, respectively. (c,d) Normalized accuracy retention for Task A and Task B, respectively. The solid orange lines and dashed green lines represent the normalized accuracy retention computed from the average accuracy over 50 trials for IQFMs (with contrastive learning) and QCNN, respectively.}\label{fig:quantum:noise}
\end{figure}

Another remarkable advantage of IQFMs is its robustness against noise. To verify this, we evaluate the performance of IQFMs (with contrastive learning) under two types of noise: physical RX noise and statistical errors due to a limited number of measurement shots.
For physical RX noise, we apply random RX rotations with angles in the range $[0, 2\pi p]$ to the data, where $p$ represents the noise level. 
In this setting, expectation values of observables are computed exactly, corresponding to an infinite number of measurement shots.
Figure~\ref{fig:quantum:noise} presents an analysis of the noise robustness of IQFMs with $L=5$ layers using contrastive learning compared to QCNN across various noise levels. While the test accuracy of both models decreases as the noise level $p$ increases, absolute accuracy alone does not fully characterize robustness against noise. To assess noise robustness beyond absolute accuracy, we therefore analyze the relative performance degradation using a normalized accuracy retention metric $R(p)=\text{Acc}(p)/\text{Acc}(0)$ ($\text{Acc}(p)$ denotes accuracy for the noise level $p$). Figure~\ref{fig:quantum:noise}(a,b) report the test accuracy of IQFMs and QCNN as a function of the noise level $p$, averaged over 50 independent trials for Tasks A and B, respectively. Based on these averaged accuracies, we compute the normalized accuracy retention shown in Fig.~\ref{fig:quantum:noise}(c,d), which is used to illustrate the relative degradation trend under increasing noise.

We observe that IQFMs exhibit superior noise robustness across a wide range of noise levels. For Task A, except at low noise levels ($p = 0.05$ and 0.1), IQFMs consistently achieve higher accuracy retention than QCNN by up to 18\%. Moreover, for Task B, IQFMs exhibit higher accuracy retention than QCNN by up to 27\% across all investigated noise levels.

\begin{figure}
		\includegraphics[width=8.7cm]{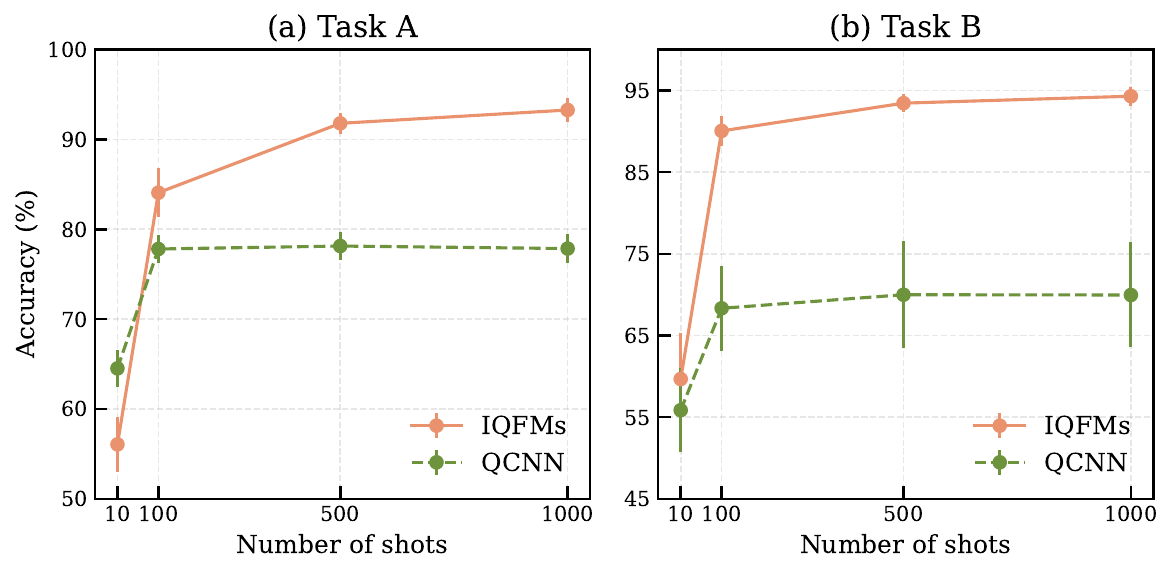}
          \protect\caption{Test accuracy for (a) Task A and (b) Task B of IQFMs and QCNN across different numbers of measurement shots. The solid orange lines and dashed green lines represent the average accuracy over 50 trials (with error bars) for IQFMs (with contrastive learning) and QCNN, respectively.}\label{fig:quantum:finite_shot}
\end{figure}

We also investigate the impact of statistical errors arising from a limited number of measurement shots. We conduct numerical experiments under finite-shot conditions, where expectation values of observables are estimated via sampling.
In our finite-shot experiments, the expectation values are estimated from $N_{\text{shot}}$ measurement shots per basis. The resulting finite-shot feature vector, denoted by $\hat{\bg}_l$, is used in the classical augmentation as $\bh_l=\cA_l(\bW_l, \hat{\bg}_l)$. During representation learning, shot samples are regenerated at each outer epoch and kept fixed across the inner epochs, reflecting the separation between quantum data collection (outer epoch) and purely classical optimization (inner epoch). Unless otherwise stated, all other results use exact expectation values (infinite-shot limit).
Figure~\ref{fig:quantum:finite_shot} shows the test accuracy for Task A and B, evaluated over 50 experimental runs with different random parameter initializations for both IQFMs with $L=5$ layers using contrastive learning and QCNN across various numbers of shots. The results indicate that the IQFMs framework achieves significantly higher test accuracy than QCNN as the number of shots increases, even though a similar number of training epochs is considered for both methods.
These findings suggest that the IQFMs framework is more robust to both physical and statistical noise, making it a reliable choice for noisy real-world quantum data.

Besides the variational QCNN baseline considered above, Ref.~\cite{cong2019quantum} provides an analytically constructed exact QCNN circuit specifically designed for Task A. In Appendix A, we further benchmark IQFMs against the exact QCNN under finite-shot measurements.  We find that the exact QCNN is advantageous in the very low-shot regime (10-100 shots), whereas IQFMs surpasses it once the number of shots becomes moderate-to-high (500-1,000 shots).

Finally, we note that IQFMs share an important conceptual connection with classical shadow–based learning approaches. In both frameworks, expressive features are extracted from quantum states via measurements in multiple bases. However, unlike classical shadow methods, IQFMs incorporate adaptive structures to broaden their applicability. In Appendix B, we further discuss the potential advantages of IQFMs and present an additional benchmark comparing IQFMs with a shadow kernel method~\cite{huang:2022:provably, Chinzei2025-wj}.

\subsection{Classical Data Classification}

We demonstrate the effectiveness of IQFMs in classical data classification tasks, showing that it achieves performance comparable to classical NNs with similar architectures. For evaluation, we use the Fashion-MNIST benchmark~\cite{fashion-mnist}, a widely used dataset for machine learning,
consisting of 60,000 training and 10,000 test samples of $28\times 28$ gray images of clothing items, each labeled with one of 10 classes.
In this demonstration, we perform 10-class classification using 5,000 randomly selected training samples and 10,000 test samples for each run.
The choice of 5,000 training samples per run reflects a practical trade-off between computational cost and sufficient per-class coverage. The full test datasets of 10,000 images are used to obtain statistically stable accuracy estimates. All reported results are averaged over 50 independent runs with different random initializations and randomly sampled training subsets, and the error bars indicate the standard deviation.

In the contrastive learning setup, the augmented vector $\bh_l(\bx)$ of the input sample $\bx$ is set as the anchor vector $\bp_l$. The positive sample $\bx^{+}$ is randomly selected from the training samples with the same label as $\bx$ and rotated 90 degrees, utilizing data augmentation. The details of data augmentation are provided in Appendix C. The negative sample $\bx^{-}$ is randomly selected from the training samples with a different label from $\bx$.

To efficiently handle relatively large data, we adopt the modular IQFMs architecture, where each module consists of $\tfrac{M}{16}$ QFMs, and each QFM processes the 16-dimensional segment of classical data.
Initially, the raw image data $\bh_{\rm raw}\in\mathbb{R}^{28\times28}=\mathbb{R}^{784}$, is linearly projected into an $M$-dimensional vector via $\bh_0 = \bW_0 \bh_{\rm raw}$, with $\bW_0 \in \mathbb{R}^{M \times 784}$.
The resulting vector $\bh_0$ is partitioned into $\tfrac{M}{16}$ blocks of 16 dimensions, each fed into a different four-qubit QFM in the first module.
Each QFM in $l$-th layer is initialized in the state $\ket{\psi}_l = \ket{0}^{\otimes 4}$ and implements the embedding circuit $\cU_{\Psi(\bh_{l-1})}$ (as shown in Fig.~\ref{fig:IQFM:quantum_data}), excluding the preprocessing circuit $P_l$. Quantum features are extracted via four Pauli-$Z$ observables, $O_i = Z_i$ ($i = 1, \dots, 4$), measured in four different bases. This yields a $d_g = 4 \times 4 \times \tfrac{M}{16} = M$-dimensional feature vector $\bg_l$. A classical augmentation step is then applied: $\bh_l = \pi \boldsymbol{\tanh}(\bW_l \bg_l / 2) \in \mathbb{R}^M$, where $\bW_l \in \mathbb{R}^{M \times M}$.
After the representation learning obtains $(\bh_1,\cdots,\bh_L)$, a three-layer feedforward NN $\mathcal{N}$ is trained to classify them into image classes.

For comparison, we also evaluate a classical NN with the same width $M$ and depth $L$, in which the QFM modules are replaced by classical nonlinear activations. Specifically, each layer is defined by $\bh_l = \boldsymbol{\tanh}(\bg_l)$ with $\bg_l = \bW_l \bh_{l-1}$. The representation learning is performed using this classical NN to produce the final feature vectors, which are then passed to the same downstream classifier $\mathcal{N}$ for evaluation.

\begin{figure}
		\includegraphics[width=8.5cm]{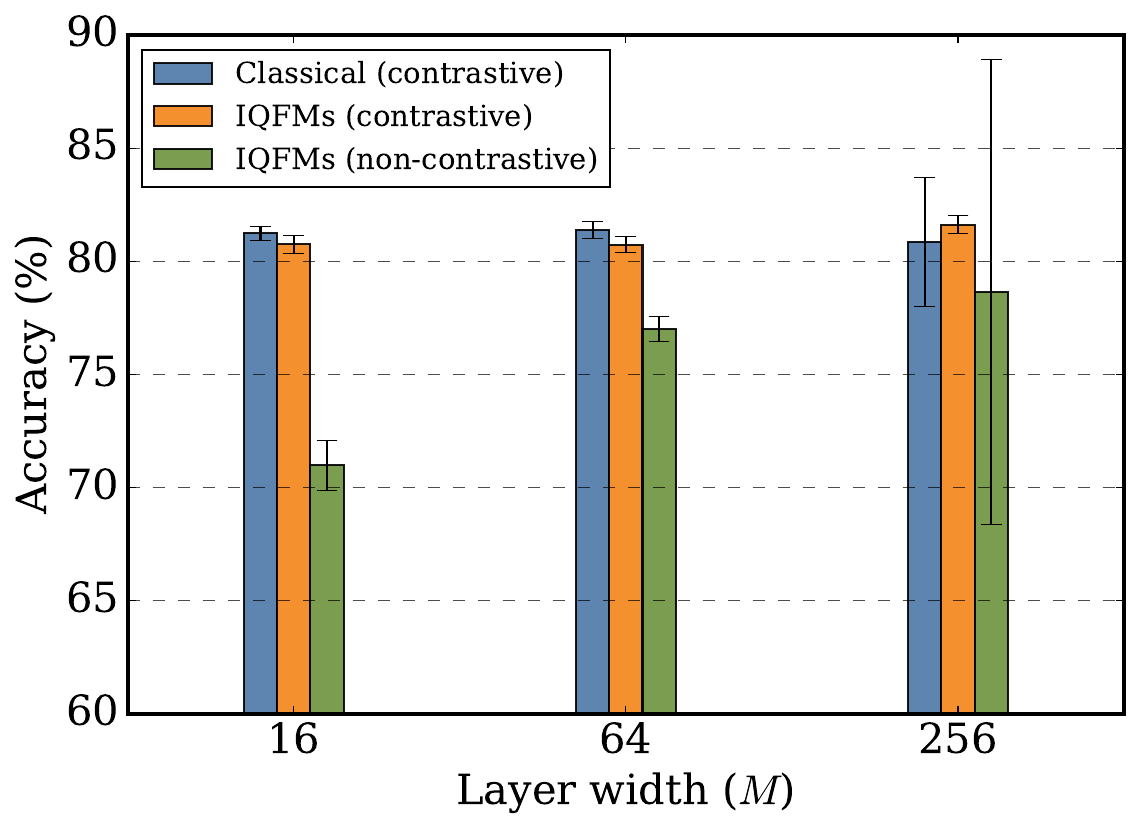}
          \protect\caption{Bar plot comparing the average test accuracy of the Fashion-MNIST classification as a function of layer width $M$ for three models: the classical NN (teal blue), IQFMs with contrastive learning (orange), and IQFMs without contrastive learning (green). Each bar represents the mean accuracy over 50 independent runs, and the error bars indicate the standard deviation.}\label{fig:MNIST}
\end{figure}

Figure~\ref{fig:MNIST} presents the test accuracy of classical neural networks and IQFMs models with $L=4$ layers and layer widths $M = 16$, $64$, and $256$, evaluated over 50 independent runs with different random parameter initializations. In the IQFMs results, a consistent improvement in test accuracy is observed as the layer width $M$ increases. The results also underscore the benefit of contrastive learning: the IQFMs framework with contrastive learning consistently outperforms its non-contrastive counterpart. Although classical NNs are well-suited to classical image data, well-designed IQFMs can achieve comparable performance to its classical counterpart with similar architecture. 

For context, other quantum classifiers have reported Fashion-MNIST classification results under different experimental settings. Zhao et al.~\cite{Zhao:2025:fashion-mnist} employ a hybrid quantum-classical classifier with relatively small training and test datasets (1,500/300), achieving a classification accuracy of 94.10\%. In contrast, Dilip et al.~\cite{Dilip:2022:fashion-mnist} benchmark data-compression-based quantum classifiers on the full Fashion-MNIST training and test datasets (60,000/10,000), reporting classification accuracies of around 80\%. Although these works use distinct circuit architectures and different dataset sizes, they provide useful reference points for situating the classification accuracy achieved by our IQFMs. While IQFMs do not reach the level of Zhao et al.~\cite{Zhao:2025:fashion-mnist}, their performance is comparable to the results of Dilip et al.~\cite{Dilip:2022:fashion-mnist}. Notably, IQFMs obtain this accuracy with fixed, non‑variational feature maps and lightweight classical training, offering a resource‑efficient option that complements existing approaches.

Combined with results on quantum data and classical data classification tasks, our findings highlight the versatility of the proposed method across both classical and quantum domains.
We note that the Fashion-MNIST experiment is not intended to claim that quantumness is necessary for this classical benchmark. Rather, it serves as a controlled sanity check demonstrating that the proposed layer-wise representation learning and the modular IQFMs architecture scale to larger classical datasets, and that inserting fixed QFM modules does not necessarily degrade performance relative to a classical network.
As an alternative to the two-step learning approach described in Fig.~\ref{fig:IQFM:overview}, there is a one-step learning method that employs contrastive learning using input samples that integrate both original data and labels; the details and results of this method are provided in Appendix F.

\section{Conclusion and Discussion}
We have introduced the IQFMs, a hybrid quantum-classical framework designed to reduce the quantum resources required for learning compared to VQAs. This framework features a deep structure that iteratively connects shallow QFMs. By training only the classical augmentation weights between QFMs, rather than the variational parameters of the quantum circuits, this design significantly reduces quantum computational runtime.

The use of contrastive learning to train the classical augmentation weights helps mitigate the effects of noise in the system. Numerical experiments demonstrate that the IQFMs framework achieves superior test accuracy compared to QCNN in tasks involving the classification of noisy quantum data. Even in classical image classification tasks, the IQFMs framework delivers performance comparable to similarly structured classical NNs. This approach offers a promising pathway to overcome current limitations and fully harness the potential of quantum-enhanced machine learning methods. Furthermore, contrastive learning is not limited to classification tasks; it can also be extended to regression tasks~\cite{zha:2023:contrast:regression}, expanding its range and effectiveness for real-world applications. 
Additionally, IQFMs could be extended with an observable-tunable readout by learning weights that form linear combinations of a large set of base observables~\cite{shen:2025:OT-EVS}. By restricting the set of observables to shadow-frugal ones (e.g., low-locality Pauli strings) and collecting a fixed set of classical shadows snapshots of the QFM output states, one could decouple quantum measurements from optimization and train the observable weights (e.g., with contrastive learning) without further quantum measurements. This extension may enhance representational expressivity while shifting more of the readout optimization to the classical side.

We also note that the IQFMs framework may naturally extend beyond static classification or regression tasks to temporal quantum information processing. In particular, while the present work used the same quantum state $\kpsi$ through the re-input structure, one could instead consider a time-indexed sequence of quantum states $\ket{\psi_t}$. This would enable IQFMs to naturally extend to quantum temporal information processing, as in quantum reservoir computing~\cite{tran:2021:temporal}. Although we focused on representation learning for static tasks in this work, exploring classical representation learning strategies within quantum temporal processing remains an intriguing direction for future research.

An alternative biologically inspired training algorithm, direct feedback alignment (DFA)~\cite{nokland:2016:dfa}, can be considered for training the classical augmentation weights in the IQFMs. DFA leverages random projections combined with alternative nonlinear activation functions, particularly when the gradient of the primary activation function is difficult to evaluate. To adapt this approach to the quantum context, various activation functions can be selected, provided the correlation between these functions and the QFM output (treated as an activation) is not close to zero~\cite{nakajima:2022:physical-4f4}, offering a flexible and potentially robust training strategy.

The simplicity of back-propagation has established it as the standard method for training classical NNs. Since back-propagation requires storing intermediate activations during the forward pass to compute gradients in the backward pass, it is incompatible with learning quantum circuits, where such intermediate states are generally inaccessible~\cite{Abbas2023-hy, Chinzei2025-pa}. Without back-propagation, the IQFMs framework relies on learning hierarchical intermediate representations that build upon one another, a concept rooted in the fundamental principles of deep learning. This enables the training of deep architectures to learn complex, hierarchical representations. One might question whether representation learning is essential. For instance, IQFMs can be redesigned to allow user-defined representations~\cite{li:2025:noprop}, offering more flexibility in its application.
Furthermore, IQFMs can be utilized in a quantum transfer learning scheme~\cite{khatun:2025:QTL}, which integrates a pre-trained classical network with a quantum classifier, providing a broader practical instantiation.

While established quantum algorithms such as the Harrow-Hassidim-Lloyd (HHL)~\cite{harrow:2009:prl} algorithm for solving linear systems and the Quantum Approximate Optimization Algorithm (QAOA)~\cite{farhi:2014:qaoa} for combinatorial optimization provide powerful frameworks with potential exponential speedups in their respective domains, they are constrained by requirements like fault-tolerant hardware, well-conditioned inputs, or deep variational circuits that can exacerbate noise on near-term devices. In contrast, the IQFMs framework emerges as an alternative powerful method that circumvents these limitations through its iterative use of shallow quantum feature maps and classical training, enhancing feasibility for noisy intermediate-scale quantum (NISQ) implementations. The IQFMs framework stands as a promising candidate for demonstrating QML applications on actual quantum devices for both classical and quantum data.

Finally, we discuss the potential for quantum advantages in IQFMs.
The first point concerns the classical simulability of the learning tasks considered here. Our benchmarks of quantum phase recognition are known to be ``locally easy'' tasks~\cite{bermejo:2024:qcnn:sim}. Once classical representations of quantum states, such as classical shadows, are obtained, these tasks can be solved efficiently by classical simulation. Consequently, exponential quantum advantages from QML algorithms are not expected in these regimes. Importantly, our aim here is not to claim quantum speedups on these benchmarks, including classical image classification, but rather to demonstrate a NISQ-friendly learning framework that avoids variational training and uses quantum hardware only for forward feature extraction. This approach thereby reduces the resource demands and instability associated with gradient estimation and noise. At the same time, we anticipate that beyond these ``locally easy'' tasks, the iterative structure of IQFMs may facilitate learning on classically nontrivial datasets, such as quantum states exhibiting nonlocal correlations or high entanglement, where classical representations become inefficient. In such regimes, IQFMs may potentially offer advantages over purely classical approaches.

The second point concerns the relationship between trainability and classical simulability in quantum models. Simple circuits that avoid exponential concentration of physical expectation values are typically easy to train, and that the functions of machine learning models based on measurement expectations tend to be classically simulable. In contrast, more complex circuits tend to be challenging to train in practice, particularly when their expressive power goes beyond what can be efficiently captured by classically simulable, expectation‑value‑based machine‑learning models. One possible way to circumvent this limitation is to move beyond physical expectation values and instead propagate measurement outcome distributions to subsequent layers. Notably, for certain circuit families, specific expectation values can be computed efficiently on a classical computer, whereas sampling from the full measurement outcome distribution is believed to be classically intractable under standard complexity-theoretic assumptions~\cite{terhal:2002:adaptive}. Exploring architectures that explicitly exploit such distribution-level information may therefore offer a promising direction for future work and enable the study of regimes where classical simulability and practical trainability do not necessarily align.

\section*{Data availability}
The datasets generated and analyzed during the current study are available in the GitHub repository at: https://github.com/FujitsuResearch/Iterative-Quantum-Feature-Maps

\begin{acknowledgments}
The authors acknowledge Yuichi Kamata and Shintaro Sato for fruitful discussions.
Special thanks are extended to Yuichi Kamata for his valuable comments on the training algorithm of IQFMs.
\end{acknowledgments}

\appendix

\section{Benchmark against an exact QCNN}

Task A is a binary quantum-phase recognition problem that distinguishes the SPT phase from non-SPT phases. Ref.~\cite{cong2019quantum} presents an analytically constructed exact QCNN circuit specifically designed for this SPT setting. In response to this point, we benchmark IQFMs against the exact QCNN under finite-shot measurements.

To enable a direct implementation of the exact QCNN circuit, we consider the open boundary variant of Task A with $n=9$ qubits. Specifically, we replace the periodic boundary Hamiltonian in Eq.~\eqref{eqn:taska} by the open boundary Hamiltonian

\begin{align}\label{eqn:taska_open}
H_A^{\mathrm{open}} = -\sum_{i=1}^{n-2} Z_i X_{i+1} Z_{i+2} - h_1 \sum_{i=1}^{n} X_i - h_2 \sum_{i=1}^{n-1} X_i X_{i+1},
\end{align}
where $X_i$ and $Z_i$ are Pauli operators acting on the $i$-th qubit. We generate ground state wave functions and labels following the same procedure as in Sec.III.A, except that we use Eq.~\eqref{eqn:taska_open} and $n=9$.

As a task-specific baseline for Task A, in our numerical experiments, we implement the exact QCNN construction of Ref.~\cite{cong2019quantum}, specialized to the open-boundary system with $n=9$ qubits. The nine-qubit register is partitioned into three blocks $(0,1,2)$, $(3,4,5)$, $(6,7,8)$, and we keep the middle qubits $\{1,4,7\}$ as the ``kept'' qubits. The circuit consists of a single convolution--pooling block (depth $d=1$) followed by a fully connected (FC) layer. In the convolution layer, we apply (i) two brickwork layers of nearest-neighbor CZ gates, (ii) CZ gates between neighboring kept qubits (1--4 and 4--7), and (iii) within each three-qubit block an $X$-controlled Toffoli operation, where the outer qubits act as controls and the middle qubit is the target. After the convolution layer, we apply SWAP gates between qubits (2,3) and (5,6). The pooling layer applies Hadamard and CZ gates within each block as follows: for $i\in\{0,3,6\}$ we apply $H_i$ followed by CZ$(i,i{+}1)$, and for $i\in\{2,5,8\}$ we apply $H_i$ followed by CZ$(i,i{-}1)$. Finally, the FC layer applies CZ gates between the kept qubits (1--4 and 4--7). For readout, we measure the Pauli-$X$ observable on the central kept qubit (wire~4) and estimate $\langle X_4 \rangle$ using $S$ shots. The exact QCNN is designed such that $\langle X_4 \rangle$ saturates near 1 in the SPT phase and is suppressed near 0 in the non-SPT phases. Based on this separation, we use the midpoint value $\langle X_4 \rangle = 0.5$ as the threshold for phase determination, and predict the SPT phase if $\widehat{\langle X_4 \rangle} > 0.5$ and the non-SPT phase otherwise. For IQFMs, we use the same architecture and training pipeline as in the main text.

\begin{figure}
		\includegraphics[width=8.7cm]{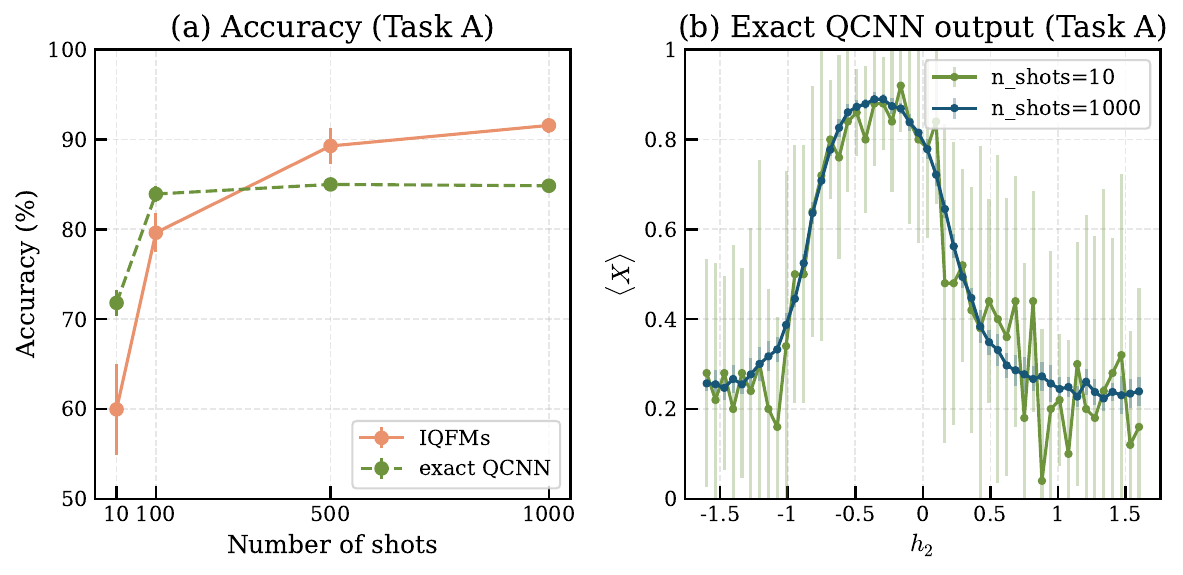}
          \protect\caption{(a) Test accuracy for Task A of IQFMs and the exact QCNN across different numbers of measurement shots. The solid orange lines and dashed green lines represent the average accuracy over 10 trials (with error bars) for IQFMs (with contrastive learning) and the exact QCNN, respectively. (b) Exact QCNN output for Task A with $h_1$ = 0.5, evaluated for different numbers of measurement shots. The solid green and blue lines indicate the average over 10 trials (with error bars) for 10 and 1,000 shots, respectively.}\label{fig:exactQCNN:shot}
\end{figure}

Figure~\ref{fig:exactQCNN:shot}(a) shows the test accuracy for IQFMs with $L=5$ layers using contrastive learning and for the exact QCNN, evaluated over 10 independent runs with different random parameters, as a function of the number of shots per measurement setting. Here, ``shots'' $S$ denotes the number of circuit repetitions per measurement setting. With this definition, the exact QCNN requires one measurement setting and thus uses $S$ circuit executions per input, whereas IQFMs employ four measurement bases per layer, resulting in a total of $4LS$ circuit executions per input (in addition to the additional circuit executions required during training).
For the IQFMs results in Fig.~\ref{fig:exactQCNN:shot}(a), we use the same dataset configuration as in Sec.III.A: the training set consists of 80 quantum states, and the test set consists of 800 quantum states. For the exact QCNN, we evaluate the model on the same test dataset of 800 quantum states as used for IQFMs, enabling a direct comparison between the two methods. The exact QCNN achieves higher accuracy in the very low-shot regime (10--100 shots), while IQFMs improves more substantially as the number of shots increases and surpasses the exact QCNN for 500 shots and above. This highlights a practical trade-off between (i) an analytic, fixed-circuit baseline that can be sample-efficient at low shots and (ii) a multi-basis feature extraction approach whose performance continues to improve with increased sampling. Here, we limit the comparison to the threshold-based approach of the exact QCNN. While its classification accuracy could potentially be improved by incorporating a more sophisticated classification model on top of the exact QCNN output, we focus here on the intrinsic decision mechanism of the exact QCNN for a direct comparison.

Figure~\ref{fig:exactQCNN:shot}(b) shows the exact QCNN output, averaged over 10 independent runs with different random parameters, as a function of the Hamiltonian parameter $h_2$, using 10 and 1,000 shots per measurement setting. For Fig.~\ref{fig:exactQCNN:shot}(b), the test dataset is constructed by fixing $h_1=0.5$ and sampling $h_2$ uniformly from the interval [−1.6,1.6] at 50 equally spaced points. Figure~\ref{fig:exactQCNN:shot}(b) shows that the measurement outcome varies systematically in the vicinity of the critical point, reflecting characteristic changes of the underlying quantum phase.

\section{Benchmark against a shadow kernel method}

IQFMs share an important conceptual connection with approaches for efficient classical snapshot acquisition from quantum data, such as classical shadows. In both frameworks, expressive features are extracted from quantum states through measurements in multiple bases. Each layer in IQFMs can be viewed as generating new measurement bases through shallow circuits, with the iterative structure enabling an adaptive‑like refinement of features. At the same time, IQFMs offer distinct advantages in integration and efficiency: unlike classical shadows, which typically involve random unitaries to estimate a broad set of observables upfront, our framework iteratively builds features tailored to the task via layer‑wise training and contrastive learning. This allows the quantum circuits to focus on extracting task‑relevant information and nonlinear features progressively, with classical augmentation weights computed to enhance feature diversity without additional quantum queries. IQFMs may therefore excel on quantum data requiring adaptive refinement or nonlinear handling, where the non‑adaptivity of classical shadows can lead to higher sample needs or lower accuracy under constraints.

Motivated by these conceptual connections and operational differences, we add a detailed comparison between IQFMs and the shadow kernel method, a representative classical shadow-based learning algorithm, using the same Task A dataset as in the main text. The shadow kernel method is a kernel-based technique built upon classical shadows, in which randomized classical shadows obtained from measurements are used to efficiently estimate the similarity (kernel value) between quantum states. The goal of this appendix is to clarify the trade-off between measurement shot efficiency and asymptotic performance under comparable quantum access costs.

IQFMs is structured as described in Sections II and III of the main text. In finite-shot experiments, the expected value is estimated by sampling with a fixed number of shots per feature for each basis and layer. The total number of shots is obtained by multiplying the number of shots per feature by the number of bases, layers, and the number of outer epochs of representation learning.

For the shadow kernel method, we employ a support vector machine (SVM) classifier with a kernel constructed from classical shadow measurements. Following the same quantum phase classification setting as for IQFMs, we consider Task A, which is a binary classification problem distinguishing the SPT phase from the other phases. We use the same dataset configuration as in Sec.III.A: the training set consists of 80 quantum states, and the test set consists of 800 quantum states. The Gram matrix $K_{ij}$ is normalized as $\tilde{K}_{ij}=K_{ij}/\sqrt{K_{ii}K_{jj}}$ to ensure numerical stability. The SVM regularization parameter $\alpha$ is selected by 5-fold cross validation on 80 training samples using a grid search over \{0.0001, 0.001, 0.01, 0.1, 1, 10, 100, 1,000, 10,000\}.

\begin{figure}
		\includegraphics[width=4.7cm]{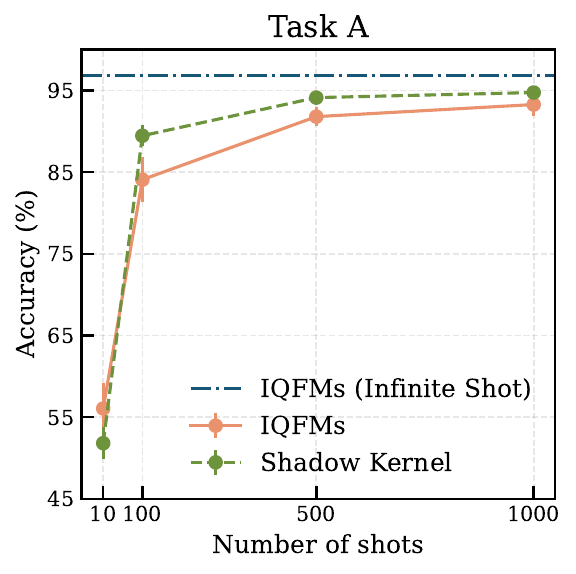}
          \protect\caption{Test accuracy for Task A of IQFMs and the shadow kernel method across different numbers of measurement shots. The solid orange lines, dashed green lines and dash-dot teal-blue lines represent the average accuracy over 50 trials for IQFMs (with contrastive learning), the shadow kernel method and IQFMs (infinite-shot), respectively.}\label{fig:shadow_kernel:shot}
\end{figure}

We compare IQFMs and the shadow kernel method in the finite-shot regime. Figure~\ref{fig:shadow_kernel:shot} reports the results for IQFMs with $L=5$ layers using contrastive learning and the shadow kernel method, evaluated over 50 independent runs with different random parameter, as a function of the number of shots per feature extraction. For both methods, accuracy improves monotonically with the number of shots, and IQFMs progressively approach their infinite-shot performance. When compared at the same number of shots, IQFMs achieve performance comparable to the shadow kernel method, with accuracy trailing by only a few percentage points. However, it is important to emphasize that in IQFMs, the horizontal axis does not correspond to the total number of measurement shots used throughout the training process.

In the IQFMs experiments, quantum feature extraction is performed using four measurement bases per layer, with five QFM layers and 100 outer epochs for representation learning. As a result, the total number of shots required by IQFMs during training is larger by approximately a factor of 2,000 compared to the shadow kernel method. In contrast, at inference time, the number of shots required by IQFMs is only about 20 times larger than that of the shadow kernel method, since inference in IQFMs uses a fixed number of expectation-value measurements per test sample.

Overall, these results highlight a trade-off between training-time and inference-time costs. Shadow kernel methods are more measurement-efficient during training, as they rely on a single data acquisition step to construct the shadow based kernel, whereas IQFMs involve iterative representation learning across multiple layers and epochs. At inference time, however, IQFMs likewise extract features for each test sample from a fixed, trained quantum circuit via a single data acquisition step, thereby avoiding the repeated measurement procedures required during training. When classical post-processing is taken into account, the overall end-to-end inference cost may favor IQFMs, since they require only a single evaluation of a trained classical predictor per test sample, independent of the training-set size. By contrast, kernel-based shadow methods typically incur a per-test overhead that scales with the number of reference states (e.g., training samples or support vectors), since kernel values must be computed classically from stored shadows between the test state and many such reference states and then aggregated. This suggests that IQFMs may offer practical advantages in settings where inference latency and total computational cost are critical, and potentially on more complex tasks benefiting from adaptive refinement or nonlinear processing.

\section{Primer on Data Augmentation and Contrastive Learning}

Some components of the IQFMs training protocol borrow standard ideas from classical representation learning, notably data augmentation and contrastive learning. This appendix provides an intuitive overview and clarifies the terminology used in the main text.

\textbf{Data augmentation.}

Data augmentation refers to label-preserving transformations applied at the level of the input data (e.g., image rotations) to generate additional views used in representation learning. Given an input example $\bx$ with label $y$, a data augmentation is a transformation $T$ such that $T(\bx)$ preserves the semantic content of $\bx$ and therefore keeps the same label $y$. Common examples in computer vision include random crops, flips, and small rotations. Such transformations increase the effective diversity of training data and encourage the learned representation to be invariant to nuisance variations. In contrastive learning, one typically forms two views of an example as $\bx^{(1)}=T_1(\bx)$ and $\bx^{(2)}=T_2(\bx)$ and treats $(\bx^{(1)},\bx^{(2)})$ as a positive pair (see, e.g., Ref.~\cite{chen:2020:simple}).

In the supervised setting used in this work, positive pairs can also be formed by selecting two different examples $(\bx_i, \bx_j)$ that share the same label ($y_i=y_j$), optionally together with label-preserving augmentations applied to either example. In our Fashion-MNIST experiment (Sec.III.B), we adopt a minimal augmentation by rotating the positive sample by 90 degrees. This encourages the learned representation to be robust to this transformation while keeping the pipeline simple and reproducible.

For quantum data, an analogous idea is to generate alternative views of the same physical class by applying transformations, such as symmetry operations, that preserve the phase label. In our quantum phase recognition experiments, positives are formed primarily by sampling different ground states from the same phase label (Sec.III.A).

\textbf{Contrastive learning and the loss used in this work.}

Contrastive learning constructs an embedding space in which representations of similar samples are close, while representations of dissimilar samples are separated by larger distance~\cite{hadsell:2006:dimensionality,gutmann:2010:contrastive,van:2018:representation,he:2020:momentum,chen:2020:simple}. Similarity is defined through a metric in representation space; in this work we use cosine similarity function, $\text{cs}(\ba, \bb)
= \frac{\ba \cdot \bb}{\|\ba\|\,\|\bb\|},$ where $\ba \cdot \bb$ is the dot product and $\|\ba\|$, $\|\bb\|$ are the vector magnitudes. 
At a given IQFMs layer $l$, the representation after classical augmentation is $\bh_l\in \bbR^{d_h}$. 
For an anchor example $\bx$, we set $\bp_l=\bh_l(\bx)$, and for a positive example $\bx^+$ (same label) and a negative example $\bx^-$ (different label), we obtain $\bh_l^+=\bh_l(\bx^+)$ and $\bh_l^-=\bh_l(\bx^-)$.

The pairwise contrastive objective adopted in the main text (Eq.~\eqref{eqn:contrast:loss}) can be written as
\begin{align}
\cC\bigl(\bh_l^{+},\bh_l^{-}\bigr)=
\log\!\biggl[1 + \exp\!\Bigl(\frac{\text{cs}(\bh_l^{-}, \bp_l) - \text{cs}(\bh_l^{+}, \bp_l)}{\tau}\Bigr)\biggr],
\end{align}
where $\tau>0$ is scale parameter.
Minimizing this cost encourages $\text{cs}(\bh_l^+, \bp_l)>\text{cs}(\bh_l^-, \bp_l)$: the loss is small when the positive is more similar to the anchor than the negative, and it increases when the negative becomes more similar than the positive.
The scale parameter $\tau$ controls how sharply the loss penalizes such violations: smaller $\tau$ makes the objective more sensitive to similarity differences, while larger $\tau$ yields smoother gradients. This objective is closely related to noise-contrastive estimation \cite{gutmann:2010:contrastive}, where negative samples act as noise examples that the model learns to distinguish from positives by optimizing a logistic-type objective over relative similarity scores.

\textbf{How we construct pairs and optimize $\bW_l$ in IQFMs.}

In IQFMs, the quantum circuit parameters $\theta_l$ are fixed and only the classical weights $\bW_l$ are trained.
For a given layer $l$, we freeze $\bW_1, \dots, \bW_{l-1}$ (trained in previous layers) and optimize only $\bW_l$ using labeled data:

\begin{enumerate}
    \item Sample an anchor example $(\bx,y)$ and compute $\bp_l=\bh_l(\bx)$.
    \item Sample a positive sample $\bx^+$ with the same label $y$ (and optionally apply a label-preserving data augmentation $T$).
    \item Sample a negative sample $\bx^-$ with the different label.
    \item Compute $\bh_l^+=\bh_l(\bx^+)$ and $\bh_l^-=\bh_l(\bx^-)$ by running the first $l$ QFM blocks to extract features and applying the classical augmentation $\cA_1,\ldots,\cA_l$ with the current weights.
    \item Update $\bW_l$ by minimizing $\cC\bigl(\bh_l^{+},\bh_l^{-}\bigr)$ with classical optimizer for a fixed number of inner steps.
\end{enumerate}

After convergence at layer $l$, $\bW_l$ is frozen and the same procedure is repeated for layer $l+1$.
All steps above apply equally when the input is a quantum state $\ket{\phi}$: one samples $(\ket{\phi}, y)$ pairs and runs the fixed QFM circuits in the corresponding states to obtain $\bg_l$ and hence $\bh_l$.

In our numerical experiments we use a single positive sample and a single negative sample per anchor to keep the required quantum data collection minimal. The framework naturally extends to multiple positive/negative samples per anchor.

\section{IQFMs in quantum data classification} 

In IQFMs model, the $l$-th QFM block includes a preprocessing circuit $P_l$ and an embedding circuit $\cU_{\Psi(\bh_{l-1})}$ (Fig.~\ref{fig:IQFM:quantum_data}).  
Here, let the symbols $\alpha, \beta, \gamma, \epsilon, \zeta$ and $\eta$ denote independent random rotation-angle parameters of the preprocessing circuit, which are fixed throughout the entire training and inference stages. 
The preprocessing circuit $P_l$ with $n$ qubits is defined as:
\begin{align} \label{eqn:preprocessing_all}
P_1 &= \Bigl(\prod_{r=1}^{d_2} B_r^{(1)}\Bigr) \Bigl(\prod_{r=1}^{d_1} A_r^{(1)}\Bigr) H^{\otimes n},\nonumber \\
P_l &= \Bigl(\prod_{r=1}^{d_2} B_r^{(l)}\Bigr) H^{\otimes n}, \quad 2\le l\le L,
\end{align}
where $H^{\otimes n}$ is the Hadamard gates acting on all qubits, and $L$ denotes the total number of layers in IQFMs. 
The preprocessing circuit consists of two main components: $A_r^{(1)}$ and $B_r^{(l)}$, which are defined as follows:
\begin{align} \label{eqn:preprocessing_circuit}
A_r^{(1)} &= \exp \left(i \sum_{j=1}^n \alpha_{r, j} Z_j\otimes Z_{j+1} \right) \exp \left(i \sum_{j=1}^n \beta_{r,j} X_j \right),\nonumber \\
B_r^{(l)} &= \exp \left(i \sum_{j=1}^n \gamma_{r, j}^{(l)} Z_j\otimes Z_{j+1} \right) \exp \left(i \sum_{j=1}^n \epsilon_{r,j}^{(l)} X_j \right) \nonumber \\
&\quad \times \exp \left(i \sum_{j=1}^n \zeta_{r,j}^{(l)} Z_j \right) \exp \left(i \sum_{j=1}^n \eta_{r,j}^{(l)} X_j \right),
\end{align}
where we have assumed the periodic boundary condition.
In our numerical experiments, the depths of the layers in the preprocessing circuit are set to $d_1 = 2$ and $d_2 = 2$.

The embedding circuit $\cU_{\Psi(\bh)}$ transforms classical data $\bh$ into a quantum state. We construct an embedding quantum circuit with $n$ qubits and a depth of $d$ layers ($d=4$ in our experiments) to embed $\bh = (h_1, h_2, \dots, h_{nd}) \in \bbR^{nd} $. In a $n$-qubit feature map with depth $d$, the classical data $\bh$ is partitioned into $d$ contiguous subvectors of length $n$: $\bh^{(m)}=(h_{n(m-1)+1}, \dots, h_{nm}), \quad m=1,2,...,d$. Then, the embedding circuit is defined as:
\begin{multline} \label{eqn:embedding}
\cU_{\Psi(\bh)} = \prod_{m=1}^{d}\Biggl[ \exp \left(-i \sum_{j=1}^n \frac{h_j^{(m)} h_{j+1}^{(m)}}{2} Z_j\otimes Z_{j+1} \right) \\
\times \exp \left(-i \sum_{j=1}^n \frac{h_j^{(m)}}{2} Z_j \right) H^{\otimes n} \Biggr].
\end{multline}

After the representation learning stage, the feature vectors from all QFM layers, $(\bh_1, \dots, \bh_L)$, are concatenated and passed to a three-layer feedforward NN for classification. 
This network consists of an input layer for linear transformation, two hidden layers with ReLU activations, and an output layer with a Softmax function to produce class probabilities. The model is trained using the cross-entropy loss with the Adam optimizer~\cite{kingma:2015:adam} (learning rate: 0.001, weight decay: $1\text{e}{-4}$) over 500 epochs. 
Training is performed in mini-batches, with loss computed and gradients propagated for parameter updates. 

\section{QCNN in quantum data classification} 

The QCNN model used in our numerical experiments consists of alternating convolutional and pooling layers. The convolutional layers apply multiple quantum operations to the input qubits, consisting of sequential RX, RZ, and RX gates followed by RZZ gates between adjacent qubits. Each gate is parameterized by trainable variational parameters, which are optimized during training. These operations are repeated for a specified number of layers, denoted by var\_depth. 
Pooling layers use CNOT gates to entangle qubits and reduce the number of active qubits by half, continuing recursively until only one qubit remains.
The final output is obtained by measuring the expectation value of the Pauli-Z operator on this remaining qubit. For binary classification (Task A), model performance is evaluated using mean squared error. For multi-class classification (Task B), a fully connected classical layer is appended to map the circuit output to a vector with a number of elements equal to the number of classes. Then, cross-entropy loss is used by comparing predicted logits to integer-encoded labels.

Figures~\ref{fig:quantum:taskab} and~\ref{fig:quantum:noise} present results under the assumption of an infinite number of measurement shots. We use the Adam optimizer~\cite{kingma:2015:adam} with a learning rate of 0.001 and weight decay of $1\text{e}{-5}$, over 1,000 epochs for Task A and 5,000 epochs for Task B. Under a finite-shot setting (Fig.~\ref{fig:quantum:finite_shot}), the parameter-shift rule is employed with mini-batch stochastic gradient descent. The batch sizes are set to 4 for Task A and 5 for Task B, with learning rates of 0.001 for both tasks, and training is performed over 300 epochs for both tasks.
The early stopping procedure is employed to prevent overfitting. 
The var\_depth of QCNN is set to 4 for Task A and 32 for Task B in Fig.~\ref{fig:quantum:noise}, and  4 for both tasks in Fig.~\ref{fig:quantum:finite_shot}.

\section{IQFMs with one-step learning}

\begin{figure}
		\includegraphics[width=8.7cm]{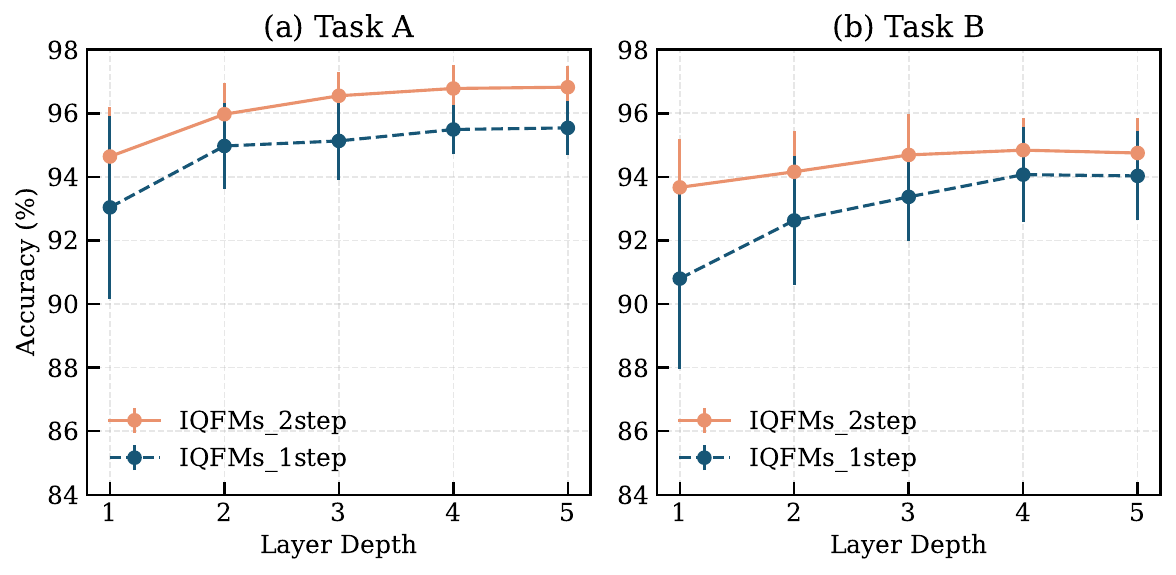}
          \protect\caption{Test accuracy for (a) Task A and (b) Task B of IQFMs with two-step learning and IQFMs with one-step learning across different layer depths. The solid orange lines and dashed teal-blue lines represent the average accuracy over 50 trials (with error bars) for IQFMs with two-step learning and IQFMs with one-step learning, respectively.}\label{fig:quantum:1step}
\end{figure}

\begin{figure}
		\includegraphics[width=8.7cm]{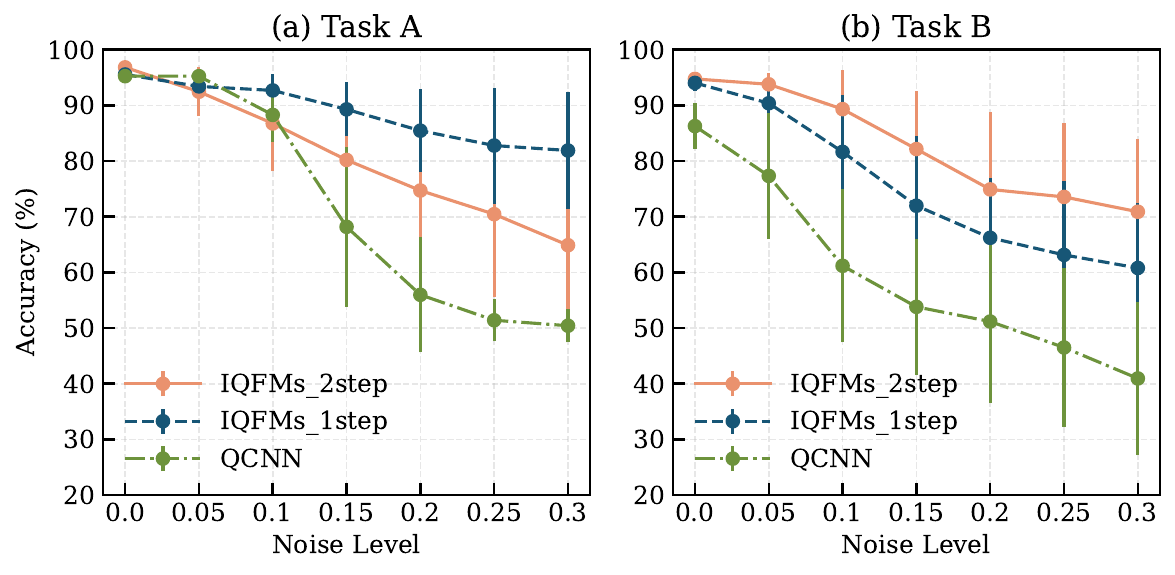}
          \protect\caption{Test accuracy for (a) Task A and (b) Task B of IQFMs with two-step learning, IQFMs with one-step learning and QCNN across different noise levels. The solid orange lines, dashed teal-blue lines and dash-dot green lines represent the average accuracy over 50 trials (with error bars) for IQFMs with two-step learning, IQFMs with one-step learning and QCNN, respectively.}\label{fig:quantum:1step_noise}
\end{figure}

We introduce an alternative training scheme for IQFMs, referred to as the one-step learning method, which employs contrastive learning using input samples that integrate both original data and labels. This method is compared with the two-step learning approach described in the main text with Fig.~\ref{fig:IQFM:overview}. In one-step learning, each input sample $\bx$ consists of a data component $\bs$ and its corresponding label $y$. A positive sample $\bx^{+} = (\bs, y)$ is paired with a negative sample $\bx^{-} = (\bs, \bar{y})$, where $\bar{y}$ denotes an incorrect label. The anchor vector $\bp_l$ is set as a fixed random vector for each layer $l$.

During training, the contrastive loss function in Eq.~\eqref{eqn:contrast:loss} is optimized using the augmented feature vectors $\bh_1(\bx^{+})$, $\bh_1(\bx^{-})$, and the anchor $\bp_l$.
In the inference phase, given a new data input $\bs$, the model predicts its label by evaluating candidate labels ${y_1, \ldots, y_K}$. For each label $y_k$, an input sample $\bx_k = (\bs, y_k)$ is formed, and a cumulative similarity score is computed across all layers as $C_k = \sum_l \text{cs}(\bh_l(\bx_k), \bp_l)$. The predicted label corresponds to the index $k$ that maximizes $C_k$.

For quantum phase recognition tasks, the input-label integration is realized by constructing a joint input state $|\phi\rangle \otimes |\text{label}\rangle$, where $|\phi\rangle$ is the input quantum data and $|\text{label}\rangle$ is a quantum encoding of the one-hot label.

Figure~\ref{fig:quantum:1step} presents the test accuracy of IQFMs using both the two-step and one-step learning methods, evaluated across varying layer depths. All other parameters follow the configurations described in Appendix D. In the absence of noise, two-step learning consistently outperforms one-step learning across different depths.

Figure~\ref{fig:quantum:1step_noise} shows the test accuracy of IQFMs ($L=5$ layers) with two-step learning, IQFMs with one-step learning, and QCNN under varying noise levels, assuming an infinite number of measurement shots. 
The var\_depth of QCNN is set to 4 for Task A and 32 for Task B. All other parameters follow the configurations described in Appendices D and E.
The results reveal that both one-step and two-step IQFMs outperform QCNN. Interestingly, the relative performance of one-step and two-step learning depends on the task and noise conditions. Under high noise levels, one-step learning outperforms for Task A, while two-step learning remains more effective for Task B.

\bibliography{main.bib}

\providecommand{\noopsort}[1]{}\providecommand{\singleletter}[1]{#1}%
\begin{thebibliography}{64}%
\makeatletter
\providecommand \@ifxundefined [1]{%
 \@ifx{#1\undefined}
}%
\providecommand \@ifnum [1]{%
 \ifnum #1\expandafter \@firstoftwo
 \else \expandafter \@secondoftwo
 \fi
}%
\providecommand \@ifx [1]{%
 \ifx #1\expandafter \@firstoftwo
 \else \expandafter \@secondoftwo
 \fi
}%
\providecommand \natexlab [1]{#1}%
\providecommand \enquote  [1]{``#1''}%
\providecommand \bibnamefont  [1]{#1}%
\providecommand \bibfnamefont [1]{#1}%
\providecommand \citenamefont [1]{#1}%
\providecommand \href@noop [0]{\@secondoftwo}%
\providecommand \href [0]{\begingroup \@sanitize@url \@href}%
\providecommand \@href[1]{\@@startlink{#1}\@@href}%
\providecommand \@@href[1]{\endgroup#1\@@endlink}%
\providecommand \@sanitize@url [0]{\catcode `\\12\catcode `\$12\catcode `\&12\catcode `\#12\catcode `\^12\catcode `\_12\catcode `\%12\relax}%
\providecommand \@@startlink[1]{}%
\providecommand \@@endlink[0]{}%
\providecommand \url  [0]{\begingroup\@sanitize@url \@url }%
\providecommand \@url [1]{\endgroup\@href {#1}{\urlprefix }}%
\providecommand \urlprefix  [0]{URL }%
\providecommand \Eprint [0]{\href }%
\providecommand \doibase [0]{https://doi.org/}%
\providecommand \selectlanguage [0]{\@gobble}%
\providecommand \bibinfo  [0]{\@secondoftwo}%
\providecommand \bibfield  [0]{\@secondoftwo}%
\providecommand \translation [1]{[#1]}%
\providecommand \BibitemOpen [0]{}%
\providecommand \bibitemStop [0]{}%
\providecommand \bibitemNoStop [0]{.\EOS\space}%
\providecommand \EOS [0]{\spacefactor3000\relax}%
\providecommand \BibitemShut  [1]{\csname bibitem#1\endcsname}%
\let\auto@bib@innerbib\@empty
\bibitem [{\citenamefont {Havl{\'{\i}}{\v{c}}ek}\ \emph {et~al.}(2019)\citenamefont {Havl{\'{\i}}{\v{c}}ek}, \citenamefont {C{\'{o}}rcoles}, \citenamefont {Temme}, \citenamefont {Harrow}, \citenamefont {Kandala}, \citenamefont {Chow},\ and\ \citenamefont {Gambetta}}]{halvlicek:2019:supervised}%
  \BibitemOpen
  \bibfield  {author} {\bibinfo {author} {\bibfnamefont {V.}~\bibnamefont {Havl{\'{\i}}{\v{c}}ek}}, \bibinfo {author} {\bibfnamefont {A.~D.}\ \bibnamefont {C{\'{o}}rcoles}}, \bibinfo {author} {\bibfnamefont {K.}~\bibnamefont {Temme}}, \bibinfo {author} {\bibfnamefont {A.~W.}\ \bibnamefont {Harrow}}, \bibinfo {author} {\bibfnamefont {A.}~\bibnamefont {Kandala}}, \bibinfo {author} {\bibfnamefont {J.~M.}\ \bibnamefont {Chow}},\ and\ \bibinfo {author} {\bibfnamefont {J.~M.}\ \bibnamefont {Gambetta}},\ }\bibfield  {title} {\bibinfo {title} {Supervised learning with quantum-enhanced feature spaces},\ }\href {https://doi.org/10.1038/s41586-019-0980-2} {\bibfield  {journal} {\bibinfo  {journal} {Nature}\ }\textbf {\bibinfo {volume} {567}},\ \bibinfo {pages} {209} (\bibinfo {year} {2019})}\BibitemShut {NoStop}%
\bibitem [{\citenamefont {Schuld}\ and\ \citenamefont {Killoran}(2019)}]{schuld:2019:feature}%
  \BibitemOpen
  \bibfield  {author} {\bibinfo {author} {\bibfnamefont {M.}~\bibnamefont {Schuld}}\ and\ \bibinfo {author} {\bibfnamefont {N.}~\bibnamefont {Killoran}},\ }\bibfield  {title} {\bibinfo {title} {Quantum machine learning in feature \text{Hilbert} spaces},\ }\href {https://doi.org/10.1103/PhysRevLett.122.040504} {\bibfield  {journal} {\bibinfo  {journal} {Phys. Rev. Lett.}\ }\textbf {\bibinfo {volume} {122}},\ \bibinfo {pages} {040504} (\bibinfo {year} {2019})}\BibitemShut {NoStop}%
\bibitem [{\citenamefont {Bravyi}\ \emph {et~al.}(2018)\citenamefont {Bravyi}, \citenamefont {Gosset},\ and\ \citenamefont {K\"{o}nig}}]{bravyi:2018:qadv}%
  \BibitemOpen
  \bibfield  {author} {\bibinfo {author} {\bibfnamefont {S.}~\bibnamefont {Bravyi}}, \bibinfo {author} {\bibfnamefont {D.}~\bibnamefont {Gosset}},\ and\ \bibinfo {author} {\bibfnamefont {R.}~\bibnamefont {K\"{o}nig}},\ }\bibfield  {title} {\bibinfo {title} {Quantum advantage with shallow circuits},\ }\href {https://doi.org/10.1126/science.aar3106} {\bibfield  {journal} {\bibinfo  {journal} {Science}\ }\textbf {\bibinfo {volume} {362}},\ \bibinfo {pages} {308} (\bibinfo {year} {2018})}\BibitemShut {NoStop}%
\bibitem [{\citenamefont {Liu}\ \emph {et~al.}(2021)\citenamefont {Liu}, \citenamefont {Arunachalam},\ and\ \citenamefont {Temme}}]{liu:2020:rigorous}%
  \BibitemOpen
  \bibfield  {author} {\bibinfo {author} {\bibfnamefont {Y.}~\bibnamefont {Liu}}, \bibinfo {author} {\bibfnamefont {S.}~\bibnamefont {Arunachalam}},\ and\ \bibinfo {author} {\bibfnamefont {K.}~\bibnamefont {Temme}},\ }\bibfield  {title} {\bibinfo {title} {A rigorous and robust quantum speed-up in supervised machine learning},\ }\href {https://doi.org/10.1038/s41567-021-01287-z} {\bibfield  {journal} {\bibinfo  {journal} {Nat. Phys.}\ }\textbf {\bibinfo {volume} {17}},\ \bibinfo {pages} {1013} (\bibinfo {year} {2021})}\BibitemShut {NoStop}%
\bibitem [{\citenamefont {Goto}\ \emph {et~al.}(2021)\citenamefont {Goto}, \citenamefont {Tran},\ and\ \citenamefont {Nakajima}}]{tran:2021:prl:uap}%
  \BibitemOpen
  \bibfield  {author} {\bibinfo {author} {\bibfnamefont {T.}~\bibnamefont {Goto}}, \bibinfo {author} {\bibfnamefont {Q.~H.}\ \bibnamefont {Tran}},\ and\ \bibinfo {author} {\bibfnamefont {K.}~\bibnamefont {Nakajima}},\ }\bibfield  {title} {\bibinfo {title} {Universal approximation property of quantum machine learning models in quantum-enhanced feature spaces},\ }\href {https://doi.org/10.1103/PhysRevLett.127.090506} {\bibfield  {journal} {\bibinfo  {journal} {Phys. Rev. Lett.}\ }\textbf {\bibinfo {volume} {127}},\ \bibinfo {pages} {090506} (\bibinfo {year} {2021})}\BibitemShut {NoStop}%
\bibitem [{\citenamefont {Huang}\ \emph {et~al.}(2021)\citenamefont {Huang}, \citenamefont {Broughton}, \citenamefont {Mohseni}, \citenamefont {Babbush}, \citenamefont {Boixo}, \citenamefont {Neven},\ and\ \citenamefont {McClean}}]{huang:2021:power}%
  \BibitemOpen
  \bibfield  {author} {\bibinfo {author} {\bibfnamefont {H.-Y.}\ \bibnamefont {Huang}}, \bibinfo {author} {\bibfnamefont {M.}~\bibnamefont {Broughton}}, \bibinfo {author} {\bibfnamefont {M.}~\bibnamefont {Mohseni}}, \bibinfo {author} {\bibfnamefont {R.}~\bibnamefont {Babbush}}, \bibinfo {author} {\bibfnamefont {S.}~\bibnamefont {Boixo}}, \bibinfo {author} {\bibfnamefont {H.}~\bibnamefont {Neven}},\ and\ \bibinfo {author} {\bibfnamefont {J.~R.}\ \bibnamefont {McClean}},\ }\bibfield  {title} {\bibinfo {title} {Power of data in quantum machine learning},\ }\href {https://doi.org/10.1038/s41467-021-22539-9} {\bibfield  {journal} {\bibinfo  {journal} {Nat. Commun.}\ }\textbf {\bibinfo {volume} {12}},\ \bibinfo {pages} {2631} (\bibinfo {year} {2021})}\BibitemShut {NoStop}%
\bibitem [{\citenamefont {Lloyd}\ \emph {et~al.}(2020)\citenamefont {Lloyd}, \citenamefont {Schuld}, \citenamefont {Ijaz}, \citenamefont {Izaac},\ and\ \citenamefont {Killoran}}]{lloyd:2020:quantum:embeddings}%
  \BibitemOpen
  \bibfield  {author} {\bibinfo {author} {\bibfnamefont {S.}~\bibnamefont {Lloyd}}, \bibinfo {author} {\bibfnamefont {M.}~\bibnamefont {Schuld}}, \bibinfo {author} {\bibfnamefont {A.}~\bibnamefont {Ijaz}}, \bibinfo {author} {\bibfnamefont {J.}~\bibnamefont {Izaac}},\ and\ \bibinfo {author} {\bibfnamefont {N.}~\bibnamefont {Killoran}},\ }\bibfield  {title} {\bibinfo {title} {{Quantum embeddings for machine learning}},\ }\Eprint {https://arxiv.org/abs/2001.03622} {arXiv:2001.03622}  (\bibinfo {year} {2020})\BibitemShut {NoStop}%
\bibitem [{\citenamefont {Rodriguez-Grasa}\ \emph {et~al.}(2025)\citenamefont {Rodriguez-Grasa}, \citenamefont {Ban},\ and\ \citenamefont {Sanz}}]{rodgriguez:2025:prr:neural:kernel}%
  \BibitemOpen
  \bibfield  {author} {\bibinfo {author} {\bibfnamefont {P.}~\bibnamefont {Rodriguez-Grasa}}, \bibinfo {author} {\bibfnamefont {Y.}~\bibnamefont {Ban}},\ and\ \bibinfo {author} {\bibfnamefont {M.}~\bibnamefont {Sanz}},\ }\bibfield  {title} {\bibinfo {title} {Neural quantum kernels: Training quantum kernels with quantum neural networks},\ }\href {https://doi.org/10.1103/xphb-x2g4} {\bibfield  {journal} {\bibinfo  {journal} {Phys. Rev. Res.}\ }\textbf {\bibinfo {volume} {7}},\ \bibinfo {pages} {023269} (\bibinfo {year} {2025})}\BibitemShut {NoStop}%
\bibitem [{\citenamefont {Mitarai}\ \emph {et~al.}(2018)\citenamefont {Mitarai}, \citenamefont {Negoro}, \citenamefont {Kitagawa},\ and\ \citenamefont {Fujii}}]{mitarai:2018:circuit}%
  \BibitemOpen
  \bibfield  {author} {\bibinfo {author} {\bibfnamefont {K.}~\bibnamefont {Mitarai}}, \bibinfo {author} {\bibfnamefont {M.}~\bibnamefont {Negoro}}, \bibinfo {author} {\bibfnamefont {M.}~\bibnamefont {Kitagawa}},\ and\ \bibinfo {author} {\bibfnamefont {K.}~\bibnamefont {Fujii}},\ }\bibfield  {title} {\bibinfo {title} {Quantum circuit learning},\ }\href {https://doi.org/10.1103/PhysRevA.98.032309} {\bibfield  {journal} {\bibinfo  {journal} {Phys. Rev. A}\ }\textbf {\bibinfo {volume} {98}},\ \bibinfo {pages} {032309} (\bibinfo {year} {2018})}\BibitemShut {NoStop}%
\bibitem [{\citenamefont {Anschuetz}\ and\ \citenamefont {Kiani}(2022)}]{ansachuetz:2022:natcom:VQA}%
  \BibitemOpen
  \bibfield  {author} {\bibinfo {author} {\bibfnamefont {E.~R.}\ \bibnamefont {Anschuetz}}\ and\ \bibinfo {author} {\bibfnamefont {B.~T.}\ \bibnamefont {Kiani}},\ }\bibfield  {title} {\bibinfo {title} {{Quantum variational algorithms are swamped with traps}},\ }\href {https://doi.org/10.1038/s41467-022-35364-5} {\bibfield  {journal} {\bibinfo  {journal} {Nat. Commun.}\ }\textbf {\bibinfo {volume} {13}},\ \bibinfo {pages} {7760} (\bibinfo {year} {2022})}\BibitemShut {NoStop}%
\bibitem [{\citenamefont {Bittel}\ and\ \citenamefont {Kliesch}(2021)}]{bittel:2021:prl:VQANP}%
  \BibitemOpen
  \bibfield  {author} {\bibinfo {author} {\bibfnamefont {L.}~\bibnamefont {Bittel}}\ and\ \bibinfo {author} {\bibfnamefont {M.}~\bibnamefont {Kliesch}},\ }\bibfield  {title} {\bibinfo {title} {Training variational quantum algorithms is np-hard},\ }\href {https://doi.org/10.1103/PhysRevLett.127.120502} {\bibfield  {journal} {\bibinfo  {journal} {Phys. Rev. Lett.}\ }\textbf {\bibinfo {volume} {127}},\ \bibinfo {pages} {120502} (\bibinfo {year} {2021})}\BibitemShut {NoStop}%
\bibitem [{\citenamefont {McClean}\ \emph {et~al.}(2018)\citenamefont {McClean}, \citenamefont {Boixo}, \citenamefont {Smelyanskiy}, \citenamefont {Babbush},\ and\ \citenamefont {Neven}}]{clean:2018:natcom:barren}%
  \BibitemOpen
  \bibfield  {author} {\bibinfo {author} {\bibfnamefont {J.~R.}\ \bibnamefont {McClean}}, \bibinfo {author} {\bibfnamefont {S.}~\bibnamefont {Boixo}}, \bibinfo {author} {\bibfnamefont {V.~N.}\ \bibnamefont {Smelyanskiy}}, \bibinfo {author} {\bibfnamefont {R.}~\bibnamefont {Babbush}},\ and\ \bibinfo {author} {\bibfnamefont {H.}~\bibnamefont {Neven}},\ }\bibfield  {title} {\bibinfo {title} {{Barren plateaus in quantum neural network training landscapes}},\ }\href {https://doi.org/10.1038/s41467-018-07090-4} {\bibfield  {journal} {\bibinfo  {journal} {Nat. Commun.}\ }\textbf {\bibinfo {volume} {9}},\ \bibinfo {pages} {4812} (\bibinfo {year} {2018})}\BibitemShut {NoStop}%
\bibitem [{\citenamefont {Cerezo}\ \emph {et~al.}(2021)\citenamefont {Cerezo}, \citenamefont {Sone}, \citenamefont {Volkoff}, \citenamefont {Cincio},\ and\ \citenamefont {Coles}}]{cerezo:2021:natcom:barren}%
  \BibitemOpen
  \bibfield  {author} {\bibinfo {author} {\bibfnamefont {M.}~\bibnamefont {Cerezo}}, \bibinfo {author} {\bibfnamefont {A.}~\bibnamefont {Sone}}, \bibinfo {author} {\bibfnamefont {T.}~\bibnamefont {Volkoff}}, \bibinfo {author} {\bibfnamefont {L.}~\bibnamefont {Cincio}},\ and\ \bibinfo {author} {\bibfnamefont {P.~J.}\ \bibnamefont {Coles}},\ }\bibfield  {title} {\bibinfo {title} {{Cost function dependent barren plateaus in shallow parametrized quantum circuits}},\ }\href {https://doi.org/10.1038/s41467-021-21728-w} {\bibfield  {journal} {\bibinfo  {journal} {Nat. Commun.}\ }\textbf {\bibinfo {volume} {12}},\ \bibinfo {pages} {1791} (\bibinfo {year} {2021})}\BibitemShut {NoStop}%
\bibitem [{\citenamefont {Ortiz~Marrero}\ \emph {et~al.}(2021)\citenamefont {Ortiz~Marrero}, \citenamefont {Kieferov\'a},\ and\ \citenamefont {Wiebe}}]{marrero:2021:prxquant:barren}%
  \BibitemOpen
  \bibfield  {author} {\bibinfo {author} {\bibfnamefont {C.}~\bibnamefont {Ortiz~Marrero}}, \bibinfo {author} {\bibfnamefont {M.}~\bibnamefont {Kieferov\'a}},\ and\ \bibinfo {author} {\bibfnamefont {N.}~\bibnamefont {Wiebe}},\ }\bibfield  {title} {\bibinfo {title} {Entanglement-induced barren plateaus},\ }\href {https://doi.org/10.1103/PRXQuantum.2.040316} {\bibfield  {journal} {\bibinfo  {journal} {PRX Quantum}\ }\textbf {\bibinfo {volume} {2}},\ \bibinfo {pages} {040316} (\bibinfo {year} {2021})}\BibitemShut {NoStop}%
\bibitem [{\citenamefont {Gil-Fuster}\ \emph {et~al.}(2024)\citenamefont {Gil-Fuster}, \citenamefont {Gyurik}, \citenamefont {Pérez-Salinas},\ and\ \citenamefont {Dunjko}}]{fuster:2024:VQA:dequan}%
  \BibitemOpen
  \bibfield  {author} {\bibinfo {author} {\bibfnamefont {E.}~\bibnamefont {Gil-Fuster}}, \bibinfo {author} {\bibfnamefont {C.}~\bibnamefont {Gyurik}}, \bibinfo {author} {\bibfnamefont {A.}~\bibnamefont {Pérez-Salinas}},\ and\ \bibinfo {author} {\bibfnamefont {V.}~\bibnamefont {Dunjko}},\ }\bibfield  {title} {\bibinfo {title} {{On the relation between trainability and dequantization of variational quantum learning models}},\ }\Eprint {https://arxiv.org/abs/2406.07072} {arXiv:2406.07072}  (\bibinfo {year} {2024})\BibitemShut {NoStop}%
\bibitem [{\citenamefont {Bermejo}\ \emph {et~al.}(2024)\citenamefont {Bermejo}, \citenamefont {Braccia}, \citenamefont {Rudolph}, \citenamefont {Holmes}, \citenamefont {Cincio},\ and\ \citenamefont {Cerezo}}]{bermejo:2024:qcnn:sim}%
  \BibitemOpen
  \bibfield  {author} {\bibinfo {author} {\bibfnamefont {P.}~\bibnamefont {Bermejo}}, \bibinfo {author} {\bibfnamefont {P.}~\bibnamefont {Braccia}}, \bibinfo {author} {\bibfnamefont {M.~S.}\ \bibnamefont {Rudolph}}, \bibinfo {author} {\bibfnamefont {Z.}~\bibnamefont {Holmes}}, \bibinfo {author} {\bibfnamefont {L.}~\bibnamefont {Cincio}},\ and\ \bibinfo {author} {\bibfnamefont {M.}~\bibnamefont {Cerezo}},\ }\bibfield  {title} {\bibinfo {title} {{Quantum Convolutional Neural Networks are Effectively Classically Simulable}},\ }\Eprint {https://arxiv.org/abs/2408.12739} {arXiv:2408.12739}  (\bibinfo {year} {2024})\BibitemShut {NoStop}%
\bibitem [{\citenamefont {Lerch}\ \emph {et~al.}(2024)\citenamefont {Lerch}, \citenamefont {Puig}, \citenamefont {Rudolph}, \citenamefont {Angrisani}, \citenamefont {Jones}, \citenamefont {Cerezo}, \citenamefont {Thanasilp},\ and\ \citenamefont {Holmes}}]{lerch:2024:quant:sim}%
  \BibitemOpen
  \bibfield  {author} {\bibinfo {author} {\bibfnamefont {S.}~\bibnamefont {Lerch}}, \bibinfo {author} {\bibfnamefont {R.}~\bibnamefont {Puig}}, \bibinfo {author} {\bibfnamefont {M.~S.}\ \bibnamefont {Rudolph}}, \bibinfo {author} {\bibfnamefont {A.}~\bibnamefont {Angrisani}}, \bibinfo {author} {\bibfnamefont {T.}~\bibnamefont {Jones}}, \bibinfo {author} {\bibfnamefont {M.}~\bibnamefont {Cerezo}}, \bibinfo {author} {\bibfnamefont {S.}~\bibnamefont {Thanasilp}},\ and\ \bibinfo {author} {\bibfnamefont {Z.}~\bibnamefont {Holmes}},\ }\bibfield  {title} {\bibinfo {title} {{Efficient quantum-enhanced classical simulation for patches of quantum landscapes}},\ }\Eprint {https://arxiv.org/abs/2411.19896} {arXiv:2411.19896}  (\bibinfo {year} {2024})\BibitemShut {NoStop}%
\bibitem [{\citenamefont {Abbas}\ \emph {et~al.}(2023)\citenamefont {Abbas}, \citenamefont {King}, \citenamefont {Huang}, \citenamefont {Huggins}, \citenamefont {Movassagh}, \citenamefont {Gilboa},\ and\ \citenamefont {McClean}}]{Abbas2023-hy}%
  \BibitemOpen
  \bibfield  {author} {\bibinfo {author} {\bibfnamefont {A.}~\bibnamefont {Abbas}}, \bibinfo {author} {\bibfnamefont {R.}~\bibnamefont {King}}, \bibinfo {author} {\bibfnamefont {H.-Y.}\ \bibnamefont {Huang}}, \bibinfo {author} {\bibfnamefont {W.~J.}\ \bibnamefont {Huggins}}, \bibinfo {author} {\bibfnamefont {R.}~\bibnamefont {Movassagh}}, \bibinfo {author} {\bibfnamefont {D.}~\bibnamefont {Gilboa}},\ and\ \bibinfo {author} {\bibfnamefont {J.}~\bibnamefont {McClean}},\ }\bibfield  {title} {\bibinfo {title} {{On quantum backpropagation, information reuse, and cheating measurement collapse}},\ }in\ \href {https://proceedings.neurips.cc/paper_files/paper/2023/file/8c3caae2f725c8e2a55ecd600563d172-Paper-Conference.pdf} {\emph {\bibinfo {booktitle} {{Advances in Neural Information Processing Systems}}}},\ Vol.~\bibinfo {volume} {36}\ (\bibinfo {year} {2023})\BibitemShut {NoStop}%
\bibitem [{\citenamefont {Chinzei}\ \emph {et~al.}(2025{\natexlab{a}})\citenamefont {Chinzei}, \citenamefont {Yamano}, \citenamefont {Tran}, \citenamefont {Endo},\ and\ \citenamefont {Oshima}}]{Chinzei2025-pa}%
  \BibitemOpen
  \bibfield  {author} {\bibinfo {author} {\bibfnamefont {K.}~\bibnamefont {Chinzei}}, \bibinfo {author} {\bibfnamefont {S.}~\bibnamefont {Yamano}}, \bibinfo {author} {\bibfnamefont {Q.~H.}\ \bibnamefont {Tran}}, \bibinfo {author} {\bibfnamefont {Y.}~\bibnamefont {Endo}},\ and\ \bibinfo {author} {\bibfnamefont {H.}~\bibnamefont {Oshima}},\ }\bibfield  {title} {\bibinfo {title} {{Trade-off between gradient measurement efficiency and expressivity in deep quantum neural networks}},\ }\href {https://doi.org/10.1038/s41534-025-01036-7} {\bibfield  {journal} {\bibinfo  {journal} {npj Quantum Inf.}\ }\textbf {\bibinfo {volume} {11}},\ \bibinfo {pages} {79} (\bibinfo {year} {2025}{\natexlab{a}})}\BibitemShut {NoStop}%
\bibitem [{\citenamefont {Liu}\ \emph {et~al.}(2023)\citenamefont {Liu}, \citenamefont {Chen}, \citenamefont {Sun}, \citenamefont {Xue}, \citenamefont {Wu},\ and\ \citenamefont {Guo}}]{liu:2023:vqa:time}%
  \BibitemOpen
  \bibfield  {author} {\bibinfo {author} {\bibfnamefont {H.-Y.}\ \bibnamefont {Liu}}, \bibinfo {author} {\bibfnamefont {Z.-Y.}\ \bibnamefont {Chen}}, \bibinfo {author} {\bibfnamefont {T.-P.}\ \bibnamefont {Sun}}, \bibinfo {author} {\bibfnamefont {C.}~\bibnamefont {Xue}}, \bibinfo {author} {\bibfnamefont {Y.-C.}\ \bibnamefont {Wu}},\ and\ \bibinfo {author} {\bibfnamefont {G.-P.}\ \bibnamefont {Guo}},\ }\bibfield  {title} {\bibinfo {title} {{Can Variational Quantum Algorithms Demonstrate Quantum Advantages? Time Really Matters}},\ }\Eprint {https://arxiv.org/abs/2307.04089} {arXiv:2307.04089}  (\bibinfo {year} {2023})\BibitemShut {NoStop}%
\bibitem [{\citenamefont {Hadsell}\ \emph {et~al.}(2006)\citenamefont {Hadsell}, \citenamefont {Chopra},\ and\ \citenamefont {LeCun}}]{hadsell:2006:dimensionality}%
  \BibitemOpen
  \bibfield  {author} {\bibinfo {author} {\bibfnamefont {R.}~\bibnamefont {Hadsell}}, \bibinfo {author} {\bibfnamefont {S.}~\bibnamefont {Chopra}},\ and\ \bibinfo {author} {\bibfnamefont {Y.}~\bibnamefont {LeCun}},\ }\bibfield  {title} {\bibinfo {title} {Dimensionality reduction by learning an invariant mapping},\ }in\ \href {https://doi.org/10.1109/CVPR.2006.100} {\emph {\bibinfo {booktitle} {Proc. IEEE Conf. Comput. Vis. Pattern Recognit. (CVPR)}}},\ Vol.~\bibinfo {volume} {2}\ (\bibinfo  {publisher} {IEEE},\ \bibinfo {year} {2006})\ pp.\ \bibinfo {pages} {1735--1742}\BibitemShut {NoStop}%
\bibitem [{\citenamefont {Gutmann}\ and\ \citenamefont {Hyvärinen}(2010)}]{gutmann:2010:contrastive}%
  \BibitemOpen
  \bibfield  {author} {\bibinfo {author} {\bibfnamefont {M.}~\bibnamefont {Gutmann}}\ and\ \bibinfo {author} {\bibfnamefont {A.}~\bibnamefont {Hyvärinen}},\ }\bibfield  {title} {\bibinfo {title} {Noise-contrastive estimation: A new estimation principle for unnormalized statistical models},\ }in\ \href {https://proceedings.mlr.press/v9/gutmann10a.html} {\emph {\bibinfo {booktitle} {Proc. 13th Int. Conf. Artif. Intell. Stat.}}},\ Vol.~\bibinfo {volume} {9}\ (\bibinfo  {publisher} {PMLR},\ \bibinfo {year} {2010})\ pp.\ \bibinfo {pages} {297--304}\BibitemShut {NoStop}%
\bibitem [{\citenamefont {van~den Oord}\ \emph {et~al.}(2018)\citenamefont {van~den Oord}, \citenamefont {Li},\ and\ \citenamefont {Vinyals}}]{van:2018:representation}%
  \BibitemOpen
  \bibfield  {author} {\bibinfo {author} {\bibfnamefont {A.}~\bibnamefont {van~den Oord}}, \bibinfo {author} {\bibfnamefont {Y.}~\bibnamefont {Li}},\ and\ \bibinfo {author} {\bibfnamefont {O.}~\bibnamefont {Vinyals}},\ }\bibfield  {title} {\bibinfo {title} {{Representation Learning with Contrastive Predictive Coding}},\ }\Eprint {https://arxiv.org/abs/1807.03748} {arXiv:1807.03748}  (\bibinfo {year} {2018})\BibitemShut {NoStop}%
\bibitem [{\citenamefont {He}\ \emph {et~al.}(2020)\citenamefont {He}, \citenamefont {Fan}, \citenamefont {Wu}, \citenamefont {Xie},\ and\ \citenamefont {Girshick}}]{he:2020:momentum}%
  \BibitemOpen
  \bibfield  {author} {\bibinfo {author} {\bibfnamefont {K.}~\bibnamefont {He}}, \bibinfo {author} {\bibfnamefont {H.}~\bibnamefont {Fan}}, \bibinfo {author} {\bibfnamefont {Y.}~\bibnamefont {Wu}}, \bibinfo {author} {\bibfnamefont {S.}~\bibnamefont {Xie}},\ and\ \bibinfo {author} {\bibfnamefont {R.}~\bibnamefont {Girshick}},\ }\bibfield  {title} {\bibinfo {title} {Momentum contrast for unsupervised visual representation learning},\ }in\ \href {https://doi.org/10.1109/CVPR42600.2020.00975} {\emph {\bibinfo {booktitle} {Proc. IEEE Conf. Comput. Vis. Pattern Recognit. (CVPR)}}}\ (\bibinfo  {publisher} {IEEE},\ \bibinfo {year} {2020})\ pp.\ \bibinfo {pages} {9729--9738}\BibitemShut {NoStop}%
\bibitem [{\citenamefont {Chen}\ \emph {et~al.}(2020)\citenamefont {Chen}, \citenamefont {Kornblith}, \citenamefont {Norouzi},\ and\ \citenamefont {Hinton}}]{chen:2020:simple}%
  \BibitemOpen
  \bibfield  {author} {\bibinfo {author} {\bibfnamefont {T.}~\bibnamefont {Chen}}, \bibinfo {author} {\bibfnamefont {S.}~\bibnamefont {Kornblith}}, \bibinfo {author} {\bibfnamefont {M.}~\bibnamefont {Norouzi}},\ and\ \bibinfo {author} {\bibfnamefont {G.}~\bibnamefont {Hinton}},\ }\bibfield  {title} {\bibinfo {title} {A simple framework for contrastive learning of visual representations},\ }in\ \href {http://proceedings.mlr.press/v119/chen20j.html} {\emph {\bibinfo {booktitle} {Proc. 37th Int. Conf. Mach. Learn. (ICML)}}},\ Vol.\ \bibinfo {volume} {119}\ (\bibinfo  {publisher} {PMLR},\ \bibinfo {year} {2020})\ pp.\ \bibinfo {pages} {1597--1607}\BibitemShut {NoStop}%
\bibitem [{\citenamefont {Cong}\ \emph {et~al.}(2019)\citenamefont {Cong}, \citenamefont {Choi},\ and\ \citenamefont {Lukin}}]{cong2019quantum}%
  \BibitemOpen
  \bibfield  {author} {\bibinfo {author} {\bibfnamefont {I.}~\bibnamefont {Cong}}, \bibinfo {author} {\bibfnamefont {S.}~\bibnamefont {Choi}},\ and\ \bibinfo {author} {\bibfnamefont {M.~D.}\ \bibnamefont {Lukin}},\ }\bibfield  {title} {\bibinfo {title} {Quantum convolutional neural networks},\ }\href {https://www.nature.com/articles/s41567-019-0648-8} {\bibfield  {journal} {\bibinfo  {journal} {Nat. Phys.}\ }\textbf {\bibinfo {volume} {15}},\ \bibinfo {pages} {1273} (\bibinfo {year} {2019})}\BibitemShut {NoStop}%
\bibitem [{\citenamefont {Chinzei}\ \emph {et~al.}(2024{\natexlab{a}})\citenamefont {Chinzei}, \citenamefont {Tran}, \citenamefont {Maruyama}, \citenamefont {Oshima},\ and\ \citenamefont {Sato}}]{Chinzei2024-nm}%
  \BibitemOpen
  \bibfield  {author} {\bibinfo {author} {\bibfnamefont {K.}~\bibnamefont {Chinzei}}, \bibinfo {author} {\bibfnamefont {Q.~H.}\ \bibnamefont {Tran}}, \bibinfo {author} {\bibfnamefont {K.}~\bibnamefont {Maruyama}}, \bibinfo {author} {\bibfnamefont {H.}~\bibnamefont {Oshima}},\ and\ \bibinfo {author} {\bibfnamefont {S.}~\bibnamefont {Sato}},\ }\bibfield  {title} {\bibinfo {title} {{Splitting and parallelizing of quantum convolutional neural networks for learning translationally symmetric data}},\ }\href {https://doi.org/10.1103/physrevresearch.6.023042} {\bibfield  {journal} {\bibinfo  {journal} {Phys. Rev. Res.}\ }\textbf {\bibinfo {volume} {6}},\ \bibinfo {pages} {023042} (\bibinfo {year} {2024}{\natexlab{a}})}\BibitemShut {NoStop}%
\bibitem [{\citenamefont {Chinzei}\ \emph {et~al.}(2024{\natexlab{b}})\citenamefont {Chinzei}, \citenamefont {Tran}, \citenamefont {Endo},\ and\ \citenamefont {Oshima}}]{Chinzei2024-tl}%
  \BibitemOpen
  \bibfield  {author} {\bibinfo {author} {\bibfnamefont {K.}~\bibnamefont {Chinzei}}, \bibinfo {author} {\bibfnamefont {Q.~H.}\ \bibnamefont {Tran}}, \bibinfo {author} {\bibfnamefont {Y.}~\bibnamefont {Endo}},\ and\ \bibinfo {author} {\bibfnamefont {H.}~\bibnamefont {Oshima}},\ }\bibfield  {title} {\bibinfo {title} {{Resource-efficient equivariant quantum convolutional neural networks}},\ }\Eprint {https://arxiv.org/abs/2410.01252} {arXiv:2410.01252}  (\bibinfo {year} {2024}{\natexlab{b}})\BibitemShut {NoStop}%
\bibitem [{\citenamefont {Pérez-Salinas}\ \emph {et~al.}(2020)\citenamefont {Pérez-Salinas}, \citenamefont {Cervera-Lierta}, \citenamefont {Gil-Fuster},\ and\ \citenamefont {Latorre}}]{perez:2020:reupload}%
  \BibitemOpen
  \bibfield  {author} {\bibinfo {author} {\bibfnamefont {A.}~\bibnamefont {Pérez-Salinas}}, \bibinfo {author} {\bibfnamefont {A.}~\bibnamefont {Cervera-Lierta}}, \bibinfo {author} {\bibfnamefont {E.}~\bibnamefont {Gil-Fuster}},\ and\ \bibinfo {author} {\bibfnamefont {J.~I.}\ \bibnamefont {Latorre}},\ }\bibfield  {title} {\bibinfo {title} {{Data re-uploading for a universal quantum classifier}},\ }\href {https://doi.org/10.22331/q-2020-02-06-226} {\bibfield  {journal} {\bibinfo  {journal} {Quantum}\ }\textbf {\bibinfo {volume} {4}},\ \bibinfo {pages} {226} (\bibinfo {year} {2020})}\BibitemShut {NoStop}%
\bibitem [{\citenamefont {Gao}\ \emph {et~al.}(2022)\citenamefont {Gao}, \citenamefont {Anschuetz}, \citenamefont {Wang}, \citenamefont {Cirac},\ and\ \citenamefont {Lukin}}]{gao:2022:prx:corr}%
  \BibitemOpen
  \bibfield  {author} {\bibinfo {author} {\bibfnamefont {X.}~\bibnamefont {Gao}}, \bibinfo {author} {\bibfnamefont {E.~R.}\ \bibnamefont {Anschuetz}}, \bibinfo {author} {\bibfnamefont {S.-T.}\ \bibnamefont {Wang}}, \bibinfo {author} {\bibfnamefont {J.~I.}\ \bibnamefont {Cirac}},\ and\ \bibinfo {author} {\bibfnamefont {M.~D.}\ \bibnamefont {Lukin}},\ }\bibfield  {title} {\bibinfo {title} {Enhancing generative models via quantum correlations},\ }\href {https://doi.org/10.1103/PhysRevX.12.021037} {\bibfield  {journal} {\bibinfo  {journal} {Phys. Rev. X}\ }\textbf {\bibinfo {volume} {12}},\ \bibinfo {pages} {021037} (\bibinfo {year} {2022})}\BibitemShut {NoStop}%
\bibitem [{\citenamefont {Rudolph}\ \emph {et~al.}(2022)\citenamefont {Rudolph}, \citenamefont {Toussaint}, \citenamefont {Katabarwa}, \citenamefont {Johri}, \citenamefont {Peropadre},\ and\ \citenamefont {Perdomo-Ortiz}}]{rudolph:2022:prx:gen}%
  \BibitemOpen
  \bibfield  {author} {\bibinfo {author} {\bibfnamefont {M.~S.}\ \bibnamefont {Rudolph}}, \bibinfo {author} {\bibfnamefont {N.~B.}\ \bibnamefont {Toussaint}}, \bibinfo {author} {\bibfnamefont {A.}~\bibnamefont {Katabarwa}}, \bibinfo {author} {\bibfnamefont {S.}~\bibnamefont {Johri}}, \bibinfo {author} {\bibfnamefont {B.}~\bibnamefont {Peropadre}},\ and\ \bibinfo {author} {\bibfnamefont {A.}~\bibnamefont {Perdomo-Ortiz}},\ }\bibfield  {title} {\bibinfo {title} {Generation of high-resolution handwritten digits with an ion-trap quantum computer},\ }\href {https://doi.org/10.1103/PhysRevX.12.031010} {\bibfield  {journal} {\bibinfo  {journal} {Phys. Rev. X}\ }\textbf {\bibinfo {volume} {12}},\ \bibinfo {pages} {031010} (\bibinfo {year} {2022})}\BibitemShut {NoStop}%
\bibitem [{\citenamefont {Van Den~Nest}(2011)}]{nest:2011:expectation}%
  \BibitemOpen
  \bibfield  {author} {\bibinfo {author} {\bibfnamefont {M.}~\bibnamefont {Van Den~Nest}},\ }\bibfield  {title} {\bibinfo {title} {Simulating quantum computers with probabilistic methods},\ }\href {https://www.rintonpress.com/journals/doi/QIC11.9-10-5.html} {\bibfield  {journal} {\bibinfo  {journal} {Quantum Info. Comput.}\ }\textbf {\bibinfo {volume} {11}},\ \bibinfo {pages} {784} (\bibinfo {year} {2011})}\BibitemShut {NoStop}%
\bibitem [{\citenamefont {Xiong}\ \emph {et~al.}(2025)\citenamefont {Xiong}, \citenamefont {Facelli}, \citenamefont {Sahebi}, \citenamefont {Agnel}, \citenamefont {Chotibut}, \citenamefont {Thanasilp},\ and\ \citenamefont {Holmes}}]{xiong:2025:fundamental}%
  \BibitemOpen
  \bibfield  {author} {\bibinfo {author} {\bibfnamefont {W.}~\bibnamefont {Xiong}}, \bibinfo {author} {\bibfnamefont {G.}~\bibnamefont {Facelli}}, \bibinfo {author} {\bibfnamefont {M.}~\bibnamefont {Sahebi}}, \bibinfo {author} {\bibfnamefont {O.}~\bibnamefont {Agnel}}, \bibinfo {author} {\bibfnamefont {T.}~\bibnamefont {Chotibut}}, \bibinfo {author} {\bibfnamefont {S.}~\bibnamefont {Thanasilp}},\ and\ \bibinfo {author} {\bibfnamefont {Z.}~\bibnamefont {Holmes}},\ }\bibfield  {title} {\bibinfo {title} {On fundamental aspects of quantum extreme learning machines},\ }\href {https://link.springer.com/article/10.1007/s42484-025-00239-7} {\bibfield  {journal} {\bibinfo  {journal} {Quantum Mach. Intell.}\ }\textbf {\bibinfo {volume} {7}},\ \bibinfo {pages} {20} (\bibinfo {year} {2025})}\BibitemShut {NoStop}%
\bibitem [{\citenamefont {Sannia}\ \emph {et~al.}(2025)\citenamefont {Sannia}, \citenamefont {Giorgi},\ and\ \citenamefont {Zambrini}}]{sannia:2025:concentration}%
  \BibitemOpen
  \bibfield  {author} {\bibinfo {author} {\bibfnamefont {A.}~\bibnamefont {Sannia}}, \bibinfo {author} {\bibfnamefont {G.~L.}\ \bibnamefont {Giorgi}},\ and\ \bibinfo {author} {\bibfnamefont {R.}~\bibnamefont {Zambrini}},\ }\bibfield  {title} {\bibinfo {title} {{Exponential concentration and symmetries in Quantum Reservoir Computing}},\ }\Eprint {https://arxiv.org/abs/2505.10062} {arXiv:2505.10062}  (\bibinfo {year} {2025})\BibitemShut {NoStop}%
\bibitem [{\citenamefont {Schuld}\ \emph {et~al.}(2019)\citenamefont {Schuld}, \citenamefont {Bergholm}, \citenamefont {Gogolin}, \citenamefont {Izaac},\ and\ \citenamefont {Killoran}}]{schuld:2019:shift}%
  \BibitemOpen
  \bibfield  {author} {\bibinfo {author} {\bibfnamefont {M.}~\bibnamefont {Schuld}}, \bibinfo {author} {\bibfnamefont {V.}~\bibnamefont {Bergholm}}, \bibinfo {author} {\bibfnamefont {C.}~\bibnamefont {Gogolin}}, \bibinfo {author} {\bibfnamefont {J.}~\bibnamefont {Izaac}},\ and\ \bibinfo {author} {\bibfnamefont {N.}~\bibnamefont {Killoran}},\ }\bibfield  {title} {\bibinfo {title} {Evaluating analytic gradients on quantum hardware},\ }\href {https://doi.org/10.1103/PhysRevA.99.032331} {\bibfield  {journal} {\bibinfo  {journal} {Phys. Rev. A}\ }\textbf {\bibinfo {volume} {99}},\ \bibinfo {pages} {032331} (\bibinfo {year} {2019})}\BibitemShut {NoStop}%
\bibitem [{\citenamefont {Huang}\ \emph {et~al.}(2006)\citenamefont {Huang}, \citenamefont {Zhu},\ and\ \citenamefont {Siew}}]{huang:2006:elm}%
  \BibitemOpen
  \bibfield  {author} {\bibinfo {author} {\bibfnamefont {G.-B.}\ \bibnamefont {Huang}}, \bibinfo {author} {\bibfnamefont {Q.-Y.}\ \bibnamefont {Zhu}},\ and\ \bibinfo {author} {\bibfnamefont {C.-K.}\ \bibnamefont {Siew}},\ }\bibfield  {title} {\bibinfo {title} {Extreme learning machine: Theory and applications},\ }\href {https://doi.org/10.1016/j.neucom.2005.12.126} {\bibfield  {journal} {\bibinfo  {journal} {Neurocomputing}\ }\textbf {\bibinfo {volume} {70}},\ \bibinfo {pages} {489} (\bibinfo {year} {2006})}\BibitemShut {NoStop}%
\bibitem [{\citenamefont {Mujal}\ \emph {et~al.}(2021)\citenamefont {Mujal}, \citenamefont {Mart{\'{\i}}nez-Pe{\~{n}}a}, \citenamefont {Nokkala}, \citenamefont {Garc{\'{\i}}a-Beni}, \citenamefont {Giorgi}, \citenamefont {Soriano},\ and\ \citenamefont {Zambrini}}]{mujal:2021:opportunities}%
  \BibitemOpen
  \bibfield  {author} {\bibinfo {author} {\bibfnamefont {P.}~\bibnamefont {Mujal}}, \bibinfo {author} {\bibfnamefont {R.}~\bibnamefont {Mart{\'{\i}}nez-Pe{\~{n}}a}}, \bibinfo {author} {\bibfnamefont {J.}~\bibnamefont {Nokkala}}, \bibinfo {author} {\bibfnamefont {J.}~\bibnamefont {Garc{\'{\i}}a-Beni}}, \bibinfo {author} {\bibfnamefont {G.~L.}\ \bibnamefont {Giorgi}}, \bibinfo {author} {\bibfnamefont {M.~C.}\ \bibnamefont {Soriano}},\ and\ \bibinfo {author} {\bibfnamefont {R.}~\bibnamefont {Zambrini}},\ }\bibfield  {title} {\bibinfo {title} {Opportunities in quantum reservoir computing and extreme learning machines},\ }\href {https://doi.org/10.1002/qute.202100027} {\bibfield  {journal} {\bibinfo  {journal} {Adv. Quantum Technol.}\ }\textbf {\bibinfo {volume} {4}},\ \bibinfo {pages} {2100027} (\bibinfo {year} {2021})}\BibitemShut {NoStop}%
\bibitem [{\citenamefont {Fujii}\ and\ \citenamefont {Nakajima}(2017)}]{fujii:2017:qrc}%
  \BibitemOpen
  \bibfield  {author} {\bibinfo {author} {\bibfnamefont {K.}~\bibnamefont {Fujii}}\ and\ \bibinfo {author} {\bibfnamefont {K.}~\bibnamefont {Nakajima}},\ }\bibfield  {title} {\bibinfo {title} {Harnessing disordered-ensemble quantum dynamics for machine learning},\ }\href {https://doi.org/10.1103/PhysRevApplied.8.024030} {\bibfield  {journal} {\bibinfo  {journal} {Phys. Rev. Applied}\ }\textbf {\bibinfo {volume} {8}},\ \bibinfo {pages} {024030} (\bibinfo {year} {2017})}\BibitemShut {NoStop}%
\bibitem [{\citenamefont {Nakajima}\ \emph {et~al.}(2019)\citenamefont {Nakajima}, \citenamefont {Fujii}, \citenamefont {Negoro}, \citenamefont {Mitarai},\ and\ \citenamefont {Kitagawa}}]{nakajima:2019:qrc}%
  \BibitemOpen
  \bibfield  {author} {\bibinfo {author} {\bibfnamefont {K.}~\bibnamefont {Nakajima}}, \bibinfo {author} {\bibfnamefont {K.}~\bibnamefont {Fujii}}, \bibinfo {author} {\bibfnamefont {M.}~\bibnamefont {Negoro}}, \bibinfo {author} {\bibfnamefont {K.}~\bibnamefont {Mitarai}},\ and\ \bibinfo {author} {\bibfnamefont {M.}~\bibnamefont {Kitagawa}},\ }\bibfield  {title} {\bibinfo {title} {Boosting computational power through spatial multiplexing in quantum reservoir computing},\ }\href {https://doi.org/10.1103/PhysRevApplied.11.034021} {\bibfield  {journal} {\bibinfo  {journal} {Phys. Rev. Applied}\ }\textbf {\bibinfo {volume} {11}},\ \bibinfo {pages} {034021} (\bibinfo {year} {2019})}\BibitemShut {NoStop}%
\bibitem [{\citenamefont {Tran}\ and\ \citenamefont {Nakajima}(2021)}]{tran:2021:temporal}%
  \BibitemOpen
  \bibfield  {author} {\bibinfo {author} {\bibfnamefont {Q.~H.}\ \bibnamefont {Tran}}\ and\ \bibinfo {author} {\bibfnamefont {K.}~\bibnamefont {Nakajima}},\ }\bibfield  {title} {\bibinfo {title} {Learning temporal quantum tomography},\ }\href {https://doi.org/10.1103/PhysRevLett.127.260401} {\bibfield  {journal} {\bibinfo  {journal} {Phys. Rev. Lett.}\ }\textbf {\bibinfo {volume} {127}},\ \bibinfo {pages} {260401} (\bibinfo {year} {2021})}\BibitemShut {NoStop}%
\bibitem [{\citenamefont {Kubota}\ \emph {et~al.}(2023)\citenamefont {Kubota}, \citenamefont {Suzuki}, \citenamefont {Kobayashi}, \citenamefont {Tran}, \citenamefont {Yamamoto},\ and\ \citenamefont {Nakajima}}]{kubota:2023:temporal}%
  \BibitemOpen
  \bibfield  {author} {\bibinfo {author} {\bibfnamefont {T.}~\bibnamefont {Kubota}}, \bibinfo {author} {\bibfnamefont {Y.}~\bibnamefont {Suzuki}}, \bibinfo {author} {\bibfnamefont {S.}~\bibnamefont {Kobayashi}}, \bibinfo {author} {\bibfnamefont {Q.~H.}\ \bibnamefont {Tran}}, \bibinfo {author} {\bibfnamefont {N.}~\bibnamefont {Yamamoto}},\ and\ \bibinfo {author} {\bibfnamefont {K.}~\bibnamefont {Nakajima}},\ }\bibfield  {title} {\bibinfo {title} {Temporal information processing induced by quantum noise},\ }\href {https://doi.org/10.1103/PhysRevResearch.5.023057} {\bibfield  {journal} {\bibinfo  {journal} {Phys. Rev. Res.}\ }\textbf {\bibinfo {volume} {5}},\ \bibinfo {pages} {023057} (\bibinfo {year} {2023})}\BibitemShut {NoStop}%
\bibitem [{\citenamefont {Tran}\ and\ \citenamefont {Nakajima}(2020)}]{tran:2020:higherorder}%
  \BibitemOpen
  \bibfield  {author} {\bibinfo {author} {\bibfnamefont {Q.~H.}\ \bibnamefont {Tran}}\ and\ \bibinfo {author} {\bibfnamefont {K.}~\bibnamefont {Nakajima}},\ }\bibfield  {title} {\bibinfo {title} {{Higher-Order Quantum Reservoir Computing}},\ }\Eprint {https://arxiv.org/abs/2006.08999} {arXiv:2006.08999}  (\bibinfo {year} {2020})\BibitemShut {NoStop}%
\bibitem [{\citenamefont {Momeni}\ \emph {et~al.}(2023)\citenamefont {Momeni}, \citenamefont {Rahmani}, \citenamefont {Malléjac}, \citenamefont {del Hougne},\ and\ \citenamefont {Fleury}}]{momeni:2023:science}%
  \BibitemOpen
  \bibfield  {author} {\bibinfo {author} {\bibfnamefont {A.}~\bibnamefont {Momeni}}, \bibinfo {author} {\bibfnamefont {B.}~\bibnamefont {Rahmani}}, \bibinfo {author} {\bibfnamefont {M.}~\bibnamefont {Malléjac}}, \bibinfo {author} {\bibfnamefont {P.}~\bibnamefont {del Hougne}},\ and\ \bibinfo {author} {\bibfnamefont {R.}~\bibnamefont {Fleury}},\ }\bibfield  {title} {\bibinfo {title} {Backpropagation-free training of deep physical neural networks},\ }\href {https://doi.org/10.1126/science.adi8474} {\bibfield  {journal} {\bibinfo  {journal} {Science}\ }\textbf {\bibinfo {volume} {382}},\ \bibinfo {pages} {1297} (\bibinfo {year} {2023})}\BibitemShut {NoStop}%
\bibitem [{\citenamefont {Wu}\ \emph {et~al.}(2024)\citenamefont {Wu}, \citenamefont {Rossi}, \citenamefont {Vicentini}, \citenamefont {Astrakhantsev}, \citenamefont {Becca}, \citenamefont {Cao}, \citenamefont {Carrasquilla}, \citenamefont {Ferrari}, \citenamefont {Georges}, \citenamefont {Hibat-Allah}, \citenamefont {Imada}, \citenamefont {Läuchli}, \citenamefont {Mazzola}, \citenamefont {Mezzacapo}, \citenamefont {Millis}, \citenamefont {Robledo~Moreno}, \citenamefont {Neupert}, \citenamefont {Nomura}, \citenamefont {Nys}, \citenamefont {Parcollet}, \citenamefont {Pohle}, \citenamefont {Romero}, \citenamefont {Schmid}, \citenamefont {Silvester}, \citenamefont {Sorella}, \citenamefont {Tocchio}, \citenamefont {Wang}, \citenamefont {White}, \citenamefont {Wietek}, \citenamefont {Yang}, \citenamefont {Yang}, \citenamefont {Zhang},\ and\ \citenamefont {Carleo}}]{Wu2024-iy}%
  \BibitemOpen
  \bibfield  {author} {\bibinfo {author} {\bibfnamefont {D.}~\bibnamefont {Wu}}, \bibinfo {author} {\bibfnamefont {R.}~\bibnamefont {Rossi}}, \bibinfo {author} {\bibfnamefont {F.}~\bibnamefont {Vicentini}}, \bibinfo {author} {\bibfnamefont {N.}~\bibnamefont {Astrakhantsev}}, \bibinfo {author} {\bibfnamefont {F.}~\bibnamefont {Becca}}, \bibinfo {author} {\bibfnamefont {X.}~\bibnamefont {Cao}}, \bibinfo {author} {\bibfnamefont {J.}~\bibnamefont {Carrasquilla}}, \bibinfo {author} {\bibfnamefont {F.}~\bibnamefont {Ferrari}}, \bibinfo {author} {\bibfnamefont {A.}~\bibnamefont {Georges}}, \bibinfo {author} {\bibfnamefont {M.}~\bibnamefont {Hibat-Allah}}, \bibinfo {author} {\bibfnamefont {M.}~\bibnamefont {Imada}}, \bibinfo {author} {\bibfnamefont {A.~M.}\ \bibnamefont {Läuchli}}, \bibinfo {author} {\bibfnamefont {G.}~\bibnamefont {Mazzola}}, \bibinfo {author} {\bibfnamefont {A.}~\bibnamefont {Mezzacapo}}, \bibinfo {author} {\bibfnamefont {A.}~\bibnamefont {Millis}}, \bibinfo {author} {\bibfnamefont
  {J.}~\bibnamefont {Robledo~Moreno}}, \bibinfo {author} {\bibfnamefont {T.}~\bibnamefont {Neupert}}, \bibinfo {author} {\bibfnamefont {Y.}~\bibnamefont {Nomura}}, \bibinfo {author} {\bibfnamefont {J.}~\bibnamefont {Nys}}, \bibinfo {author} {\bibfnamefont {O.}~\bibnamefont {Parcollet}}, \bibinfo {author} {\bibfnamefont {R.}~\bibnamefont {Pohle}}, \bibinfo {author} {\bibfnamefont {I.}~\bibnamefont {Romero}}, \bibinfo {author} {\bibfnamefont {M.}~\bibnamefont {Schmid}}, \bibinfo {author} {\bibfnamefont {J.~M.}\ \bibnamefont {Silvester}}, \bibinfo {author} {\bibfnamefont {S.}~\bibnamefont {Sorella}}, \bibinfo {author} {\bibfnamefont {L.~F.}\ \bibnamefont {Tocchio}}, \bibinfo {author} {\bibfnamefont {L.}~\bibnamefont {Wang}}, \bibinfo {author} {\bibfnamefont {S.~R.}\ \bibnamefont {White}}, \bibinfo {author} {\bibfnamefont {A.}~\bibnamefont {Wietek}}, \bibinfo {author} {\bibfnamefont {Q.}~\bibnamefont {Yang}}, \bibinfo {author} {\bibfnamefont {Y.}~\bibnamefont {Yang}}, \bibinfo {author} {\bibfnamefont
  {S.}~\bibnamefont {Zhang}},\ and\ \bibinfo {author} {\bibfnamefont {G.}~\bibnamefont {Carleo}},\ }\bibfield  {title} {\bibinfo {title} {{Variational benchmarks for quantum many-body problems}},\ }\href {https://doi.org/10.1126/science.adg9774} {\bibfield  {journal} {\bibinfo  {journal} {Science}\ }\textbf {\bibinfo {volume} {386}},\ \bibinfo {pages} {296} (\bibinfo {year} {2024})}\BibitemShut {NoStop}%
\bibitem [{\citenamefont {Verresen}\ \emph {et~al.}(2017)\citenamefont {Verresen}, \citenamefont {Moessner},\ and\ \citenamefont {Pollmann}}]{PhysRevB.96.165124}%
  \BibitemOpen
  \bibfield  {author} {\bibinfo {author} {\bibfnamefont {R.}~\bibnamefont {Verresen}}, \bibinfo {author} {\bibfnamefont {R.}~\bibnamefont {Moessner}},\ and\ \bibinfo {author} {\bibfnamefont {F.}~\bibnamefont {Pollmann}},\ }\bibfield  {title} {\bibinfo {title} {One-dimensional symmetry protected topological phases and their transitions},\ }\href {https://doi.org/10.1103/PhysRevB.96.165124} {\bibfield  {journal} {\bibinfo  {journal} {Phys. Rev. B}\ }\textbf {\bibinfo {volume} {96}},\ \bibinfo {pages} {165124} (\bibinfo {year} {2017})}\BibitemShut {NoStop}%
\bibitem [{\citenamefont {Schuch}\ \emph {et~al.}(2011)\citenamefont {Schuch}, \citenamefont {P\'erez-Garc\'{\i}a},\ and\ \citenamefont {Cirac}}]{PhysRevB.84.165139}%
  \BibitemOpen
  \bibfield  {author} {\bibinfo {author} {\bibfnamefont {N.}~\bibnamefont {Schuch}}, \bibinfo {author} {\bibfnamefont {D.}~\bibnamefont {P\'erez-Garc\'{\i}a}},\ and\ \bibinfo {author} {\bibfnamefont {I.}~\bibnamefont {Cirac}},\ }\bibfield  {title} {\bibinfo {title} {Classifying quantum phases using matrix product states and projected entangled pair states},\ }\href {https://doi.org/10.1103/PhysRevB.84.165139} {\bibfield  {journal} {\bibinfo  {journal} {Phys. Rev. B}\ }\textbf {\bibinfo {volume} {84}},\ \bibinfo {pages} {165139} (\bibinfo {year} {2011})}\BibitemShut {NoStop}%
\bibitem [{\citenamefont {Recio-Armengol}\ \emph {et~al.}(2024)\citenamefont {Recio-Armengol}, \citenamefont {Schreiber}, \citenamefont {Eisert},\ and\ \citenamefont {Bravo-Prieto}}]{recio2024learning}%
  \BibitemOpen
  \bibfield  {author} {\bibinfo {author} {\bibfnamefont {E.}~\bibnamefont {Recio-Armengol}}, \bibinfo {author} {\bibfnamefont {F.~J.}\ \bibnamefont {Schreiber}}, \bibinfo {author} {\bibfnamefont {J.}~\bibnamefont {Eisert}},\ and\ \bibinfo {author} {\bibfnamefont {C.}~\bibnamefont {Bravo-Prieto}},\ }\bibfield  {title} {\bibinfo {title} {{Learning complexity gradually in quantum machine learning models}},\ }\Eprint {https://arxiv.org/abs/2411.11954} {arXiv:2411.11954}  (\bibinfo {year} {2024})\BibitemShut {NoStop}%
\bibitem [{\citenamefont {Wang}\ \emph {et~al.}(2022)\citenamefont {Wang}, \citenamefont {Ding}, \citenamefont {Gu}, \citenamefont {Lin}, \citenamefont {Pan}, \citenamefont {Chong},\ and\ \citenamefont {Han}}]{hanruiwang:2022:quantumnas}%
  \BibitemOpen
  \bibfield  {author} {\bibinfo {author} {\bibfnamefont {H.}~\bibnamefont {Wang}}, \bibinfo {author} {\bibfnamefont {Y.}~\bibnamefont {Ding}}, \bibinfo {author} {\bibfnamefont {J.}~\bibnamefont {Gu}}, \bibinfo {author} {\bibfnamefont {Y.}~\bibnamefont {Lin}}, \bibinfo {author} {\bibfnamefont {D.~Z.}\ \bibnamefont {Pan}}, \bibinfo {author} {\bibfnamefont {F.~T.}\ \bibnamefont {Chong}},\ and\ \bibinfo {author} {\bibfnamefont {S.}~\bibnamefont {Han}},\ }\bibfield  {title} {\bibinfo {title} {Quantumnas: Noise-adaptive search for robust quantum circuits},\ }in\ \href {https://doi.org/10.1109/HPCA53966.2022.00057} {\emph {\bibinfo {booktitle} {Proc. 28th IEEE Int. Symp. High-Perform. Comput. Arch.}}}\ (\bibinfo {year} {2022})\BibitemShut {NoStop}%
\bibitem [{\citenamefont {McInnes}\ \emph {et~al.}(2018)\citenamefont {McInnes}, \citenamefont {Healy},\ and\ \citenamefont {Melville}}]{mcinnes2018umap}%
  \BibitemOpen
  \bibfield  {author} {\bibinfo {author} {\bibfnamefont {L.}~\bibnamefont {McInnes}}, \bibinfo {author} {\bibfnamefont {J.}~\bibnamefont {Healy}},\ and\ \bibinfo {author} {\bibfnamefont {J.}~\bibnamefont {Melville}},\ }\bibfield  {title} {\bibinfo {title} {{UMAP: Uniform Manifold Approximation and Projection for Dimension Reduction}},\ }\Eprint {https://arxiv.org/abs/1802.03426} {arXiv:1802.03426}  (\bibinfo {year} {2018})\BibitemShut {NoStop}%
\bibitem [{\citenamefont {Huang}\ \emph {et~al.}(2022)\citenamefont {Huang}, \citenamefont {Kueng}, \citenamefont {Torlai}, \citenamefont {Albert},\ and\ \citenamefont {Preskill}}]{huang:2022:provably}%
  \BibitemOpen
  \bibfield  {author} {\bibinfo {author} {\bibfnamefont {H.-Y.}\ \bibnamefont {Huang}}, \bibinfo {author} {\bibfnamefont {R.}~\bibnamefont {Kueng}}, \bibinfo {author} {\bibfnamefont {G.}~\bibnamefont {Torlai}}, \bibinfo {author} {\bibfnamefont {V.~V.}\ \bibnamefont {Albert}},\ and\ \bibinfo {author} {\bibfnamefont {J.}~\bibnamefont {Preskill}},\ }\bibfield  {title} {\bibinfo {title} {Provably efficient machine learning for quantum many-body problems},\ }\href {https://www.science.org/doi/10.1126/science.abk3333} {\bibfield  {journal} {\bibinfo  {journal} {Science}\ }\textbf {\bibinfo {volume} {377}},\ \bibinfo {pages} {eabk3333} (\bibinfo {year} {2022})}\BibitemShut {NoStop}%
\bibitem [{\citenamefont {Chinzei}\ \emph {et~al.}(2025{\natexlab{b}})\citenamefont {Chinzei}, \citenamefont {Tran}, \citenamefont {Matsumoto}, \citenamefont {Endo},\ and\ \citenamefont {Oshima}}]{Chinzei2025-wj}%
  \BibitemOpen
  \bibfield  {author} {\bibinfo {author} {\bibfnamefont {K.}~\bibnamefont {Chinzei}}, \bibinfo {author} {\bibfnamefont {Q.~H.}\ \bibnamefont {Tran}}, \bibinfo {author} {\bibfnamefont {N.}~\bibnamefont {Matsumoto}}, \bibinfo {author} {\bibfnamefont {Y.}~\bibnamefont {Endo}},\ and\ \bibinfo {author} {\bibfnamefont {H.}~\bibnamefont {Oshima}},\ }\bibfield  {title} {\bibinfo {title} {{Learning quantum many-body data locally: A provably scalable framework}},\ }\Eprint {https://arxiv.org/abs/2509.13705} {arXiv:2509.13705}  (\bibinfo {year} {2025}{\natexlab{b}})\BibitemShut {NoStop}%
\bibitem [{\citenamefont {Xiao}\ \emph {et~al.}(2017)\citenamefont {Xiao}, \citenamefont {Rasul},\ and\ \citenamefont {Vollgraf}}]{fashion-mnist}%
  \BibitemOpen
  \bibfield  {author} {\bibinfo {author} {\bibfnamefont {H.}~\bibnamefont {Xiao}}, \bibinfo {author} {\bibfnamefont {K.}~\bibnamefont {Rasul}},\ and\ \bibinfo {author} {\bibfnamefont {R.}~\bibnamefont {Vollgraf}},\ }\bibfield  {title} {\bibinfo {title} {{Fashion-MNIST: a Novel Image Dataset for Benchmarking Machine Learning Algorithms}},\ }\Eprint {https://arxiv.org/abs/1708.07747} {arXiv:1708.07747}  (\bibinfo {year} {2017})\BibitemShut {NoStop}%
\bibitem [{\citenamefont {Zhao}\ \emph {et~al.}(2025)\citenamefont {Zhao}, \citenamefont {Siljak}, \citenamefont {Wang}, \citenamefont {He},\ and\ \citenamefont {Wang}}]{Zhao:2025:fashion-mnist}%
  \BibitemOpen
  \bibfield  {author} {\bibinfo {author} {\bibfnamefont {R.-X.}\ \bibnamefont {Zhao}}, \bibinfo {author} {\bibfnamefont {H.}~\bibnamefont {Siljak}}, \bibinfo {author} {\bibfnamefont {S.}~\bibnamefont {Wang}}, \bibinfo {author} {\bibfnamefont {Y.}~\bibnamefont {He}},\ and\ \bibinfo {author} {\bibfnamefont {Y.}~\bibnamefont {Wang}},\ }\bibfield  {title} {\bibinfo {title} {Hqcc: A hybrid quantum-classical classifier with adaptive structure},\ }\href {https://doi.org/10.1109/LSP.2025.3606843} {\bibfield  {journal} {\bibinfo  {journal} {IEEE Signal Process. Lett.}\ }\textbf {\bibinfo {volume} {32}},\ \bibinfo {pages} {3585} (\bibinfo {year} {2025})}\BibitemShut {NoStop}%
\bibitem [{\citenamefont {Dilip}\ \emph {et~al.}(2022)\citenamefont {Dilip}, \citenamefont {Liu}, \citenamefont {Smith},\ and\ \citenamefont {Pollmann}}]{Dilip:2022:fashion-mnist}%
  \BibitemOpen
  \bibfield  {author} {\bibinfo {author} {\bibfnamefont {R.}~\bibnamefont {Dilip}}, \bibinfo {author} {\bibfnamefont {Y.-J.}\ \bibnamefont {Liu}}, \bibinfo {author} {\bibfnamefont {A.}~\bibnamefont {Smith}},\ and\ \bibinfo {author} {\bibfnamefont {F.}~\bibnamefont {Pollmann}},\ }\bibfield  {title} {\bibinfo {title} {Data compression for quantum machine learning},\ }\href {https://doi.org/10.1103/PhysRevResearch.4.043007} {\bibfield  {journal} {\bibinfo  {journal} {Phys. Rev. Res.}\ }\textbf {\bibinfo {volume} {4}},\ \bibinfo {pages} {043007} (\bibinfo {year} {2022})}\BibitemShut {NoStop}%
\bibitem [{\citenamefont {Zha}\ \emph {et~al.}(2023)\citenamefont {Zha}, \citenamefont {Cao}, \citenamefont {Son}, \citenamefont {Yang},\ and\ \citenamefont {Katabi}}]{zha:2023:contrast:regression}%
  \BibitemOpen
  \bibfield  {author} {\bibinfo {author} {\bibfnamefont {K.}~\bibnamefont {Zha}}, \bibinfo {author} {\bibfnamefont {P.}~\bibnamefont {Cao}}, \bibinfo {author} {\bibfnamefont {J.}~\bibnamefont {Son}}, \bibinfo {author} {\bibfnamefont {Y.}~\bibnamefont {Yang}},\ and\ \bibinfo {author} {\bibfnamefont {D.}~\bibnamefont {Katabi}},\ }\bibfield  {title} {\bibinfo {title} {Rank-n-contrast: Learning continuous representations for regression},\ }in\ \href {https://proceedings.neurips.cc/paper_files/paper/2023/file/39e9c5913c970e3e49c2df629daff636-Paper-Conference.pdf} {\emph {\bibinfo {booktitle} {Proc. 37th Int. Conf. on Neural Info. Process. Systems}}},\ Vol.~\bibinfo {volume} {36}\ (\bibinfo {year} {2023})\ pp.\ \bibinfo {pages} {17882--17903}\BibitemShut {NoStop}%
\bibitem [{\citenamefont {Shen}\ \emph {et~al.}(2025)\citenamefont {Shen}, \citenamefont {Kurkin}, \citenamefont {P{\'e}rez-Salinas}, \citenamefont {Shishenina}, \citenamefont {Dunjko},\ and\ \citenamefont {Wang}}]{shen:2025:OT-EVS}%
  \BibitemOpen
  \bibfield  {author} {\bibinfo {author} {\bibfnamefont {K.}~\bibnamefont {Shen}}, \bibinfo {author} {\bibfnamefont {A.}~\bibnamefont {Kurkin}}, \bibinfo {author} {\bibfnamefont {A.}~\bibnamefont {P{\'e}rez-Salinas}}, \bibinfo {author} {\bibfnamefont {E.}~\bibnamefont {Shishenina}}, \bibinfo {author} {\bibfnamefont {V.}~\bibnamefont {Dunjko}},\ and\ \bibinfo {author} {\bibfnamefont {H.}~\bibnamefont {Wang}},\ }\bibfield  {title} {\bibinfo {title} {Variational quantum generative modeling by sampling expectation values of tunable observables},\ }\href {https://doi.org/10.1038/s41534-025-01121-x} {\bibfield  {journal} {\bibinfo  {journal} {npj Quantum Information}\ }\textbf {\bibinfo {volume} {11}},\ \bibinfo {pages} {178} (\bibinfo {year} {2025})}\BibitemShut {NoStop}%
\bibitem [{\citenamefont {N\o{}kland}(2016)}]{nokland:2016:dfa}%
  \BibitemOpen
  \bibfield  {author} {\bibinfo {author} {\bibfnamefont {A.}~\bibnamefont {N\o{}kland}},\ }\bibfield  {title} {\bibinfo {title} {Direct feedback alignment provides learning in deep neural networks},\ }in\ \href {https://proceedings.neurips.cc/paper_files/paper/2016/file/d490d7b4576290fa60eb31b5fc917ad1-Paper.pdf} {\emph {\bibinfo {booktitle} {Proc. 30th Int. Conf. on Neural Info. Process. Systems}}},\ Vol.~\bibinfo {volume} {29}\ (\bibinfo {year} {2016})\BibitemShut {NoStop}%
\bibitem [{\citenamefont {Nakajima}\ \emph {et~al.}(2022)\citenamefont {Nakajima}, \citenamefont {Inoue}, \citenamefont {Tanaka}, \citenamefont {Kuniyoshi}, \citenamefont {Hashimoto},\ and\ \citenamefont {Nakajima}}]{nakajima:2022:physical-4f4}%
  \BibitemOpen
  \bibfield  {author} {\bibinfo {author} {\bibfnamefont {M.}~\bibnamefont {Nakajima}}, \bibinfo {author} {\bibfnamefont {K.}~\bibnamefont {Inoue}}, \bibinfo {author} {\bibfnamefont {K.}~\bibnamefont {Tanaka}}, \bibinfo {author} {\bibfnamefont {Y.}~\bibnamefont {Kuniyoshi}}, \bibinfo {author} {\bibfnamefont {T.}~\bibnamefont {Hashimoto}},\ and\ \bibinfo {author} {\bibfnamefont {K.}~\bibnamefont {Nakajima}},\ }\bibfield  {title} {\bibinfo {title} {Physical deep learning with biologically inspired training method: gradient-free approach for physical hardware},\ }\href {https://doi.org/10.1038/s41467-022-35216-2} {\bibfield  {journal} {\bibinfo  {journal} {Nat. Commun.}\ }\textbf {\bibinfo {volume} {13}},\ \bibinfo {pages} {7847} (\bibinfo {year} {2022})}\BibitemShut {NoStop}%
\bibitem [{\citenamefont {Li}\ \emph {et~al.}(2025)\citenamefont {Li}, \citenamefont {Teh},\ and\ \citenamefont {Pascanu}}]{li:2025:noprop}%
  \BibitemOpen
  \bibfield  {author} {\bibinfo {author} {\bibfnamefont {Q.}~\bibnamefont {Li}}, \bibinfo {author} {\bibfnamefont {Y.~W.}\ \bibnamefont {Teh}},\ and\ \bibinfo {author} {\bibfnamefont {R.}~\bibnamefont {Pascanu}},\ }\bibfield  {title} {\bibinfo {title} {{NoProp: Training Neural Networks without Full Back-propagation or Full Forward-propagation}},\ }\Eprint {https://arxiv.org/abs/2503.24322} {arXiv:2503.24322}  (\bibinfo {year} {2025})\BibitemShut {NoStop}%
\bibitem [{\citenamefont {Khatun}\ and\ \citenamefont {Usman}(2025)}]{khatun:2025:QTL}%
  \BibitemOpen
  \bibfield  {author} {\bibinfo {author} {\bibfnamefont {A.}~\bibnamefont {Khatun}}\ and\ \bibinfo {author} {\bibfnamefont {M.}~\bibnamefont {Usman}},\ }\bibfield  {title} {\bibinfo {title} {Quantum transfer learning with adversarial robustness for classification of high-resolution image datasets},\ }\href {https://advanced.onlinelibrary.wiley.com/doi/abs/10.1002/qute.202400268} {\bibfield  {journal} {\bibinfo  {journal} {Adv. Quantum Technol.}\ }\textbf {\bibinfo {volume} {8}},\ \bibinfo {pages} {2400268} (\bibinfo {year} {2025})}\BibitemShut {NoStop}%
\bibitem [{\citenamefont {Harrow}\ \emph {et~al.}(2009)\citenamefont {Harrow}, \citenamefont {Hassidim},\ and\ \citenamefont {Lloyd}}]{harrow:2009:prl}%
  \BibitemOpen
  \bibfield  {author} {\bibinfo {author} {\bibfnamefont {A.~W.}\ \bibnamefont {Harrow}}, \bibinfo {author} {\bibfnamefont {A.}~\bibnamefont {Hassidim}},\ and\ \bibinfo {author} {\bibfnamefont {S.}~\bibnamefont {Lloyd}},\ }\bibfield  {title} {\bibinfo {title} {Quantum algorithm for linear systems of equations},\ }\href {https://doi.org/10.1103/PhysRevLett.103.150502} {\bibfield  {journal} {\bibinfo  {journal} {Phys. Rev. Lett.}\ }\textbf {\bibinfo {volume} {103}},\ \bibinfo {pages} {150502} (\bibinfo {year} {2009})}\BibitemShut {NoStop}%
\bibitem [{\citenamefont {Farhi}\ \emph {et~al.}(2014)\citenamefont {Farhi}, \citenamefont {Goldstone},\ and\ \citenamefont {Gutmann}}]{farhi:2014:qaoa}%
  \BibitemOpen
  \bibfield  {author} {\bibinfo {author} {\bibfnamefont {E.}~\bibnamefont {Farhi}}, \bibinfo {author} {\bibfnamefont {J.}~\bibnamefont {Goldstone}},\ and\ \bibinfo {author} {\bibfnamefont {S.}~\bibnamefont {Gutmann}},\ }\bibfield  {title} {\bibinfo {title} {{A Quantum Approximate Optimization Algorithm}},\ }\Eprint {https://arxiv.org/abs/1411.4028} {arXiv:1411.4028}  (\bibinfo {year} {2014})\BibitemShut {NoStop}%
\bibitem [{\citenamefont {Terhal}\ and\ \citenamefont {DiVincenzo}(2002)}]{terhal:2002:adaptive}%
  \BibitemOpen
  \bibfield  {author} {\bibinfo {author} {\bibfnamefont {B.~M.}\ \bibnamefont {Terhal}}\ and\ \bibinfo {author} {\bibfnamefont {D.~P.}\ \bibnamefont {DiVincenzo}},\ }\bibfield  {title} {\bibinfo {title} {{Adaptive quantum computation, constant depth quantum circuits and Arthur-Merlin games}},\ }\Eprint {https://arxiv.org/abs/quant-ph/0205133} {arXiv:quant-ph/0205133}  (\bibinfo {year} {2002})\BibitemShut {NoStop}%
\bibitem [{\citenamefont {Kingma}\ and\ \citenamefont {Ba}(2015)}]{kingma:2015:adam}%
  \BibitemOpen
  \bibfield  {author} {\bibinfo {author} {\bibfnamefont {D.~P.}\ \bibnamefont {Kingma}}\ and\ \bibinfo {author} {\bibfnamefont {J.}~\bibnamefont {Ba}},\ }\bibfield  {title} {\bibinfo {title} {Adam: A method for stochastic optimization},\ }in\ \href {https://arxiv.org/abs/1412.6980} {\emph {\bibinfo {booktitle} {Proc. 3rd Int. Conf. on Learning Representations}}}\ (\bibinfo {year} {2015})\BibitemShut {NoStop}%
\end{thebibliography}%
\end{document}